\documentclass[%
showkeys,
preprintnumbers,
 amsmath,amssymb,
longbibliography
]{revtex4-2}
\usepackage{graphicx}
\usepackage{amsmath}
\usepackage{xcolor}
\usepackage{soul}
\usepackage{multirow}
\usepackage{psfrag}
\usepackage{latexsym}
\usepackage{amsmath}
\usepackage{multirow}
\usepackage{amssymb}
\usepackage{epsfig}
\usepackage{graphicx}
\usepackage{floatrow}
\usepackage{units}
\usepackage{array}
\usepackage{xfrac}
\usepackage{bm}
\usepackage{bigints}
\usepackage{color}
\usepackage{tikz}
\usepackage{tikzscale}
\usepackage{pgfplots}
\usepackage{pifont}
\usepackage{mathtools}
\usepackage{hyperref}
\hypersetup{
    colorlinks = true,
    linkcolor = {blue},
    citecolor  = {blue},
    filecolor = {blue},
    menucolor= {blue},
    runcolor = {blue},
    urlcolor = {blue},
    linkcolor = {blue}
}
\usetikzlibrary{decorations}
\usetikzlibrary{calc,decorations.pathreplacing}
\usepackage{flushend}
\usepackage{upgreek}
\usepackage{scrextend}
\usepackage{lipsum}
\usepackage{enumitem}

\definecolor{OliveGreen}{RGB}{0,200,0}
\definecolor{orange}{RGB}{255, 100, 0}
\definecolor{magentaa}{RGB}{170, 0, 170}

\newcommand{{\rd}}[1]{{\color{OliveGreen}#1}}

\begin{document}

\preprint{APS - Physical Review Fluids}

\title{Convection velocities and velocity coupling of outer-scaled wall-pressure fluctuations in canonical turbulent boundary layers}

\author{Rahul Deshpande$^{1,2}$}\email[Corresponding author:~]{raadeshpande@gmail.com}
\author{Abdelrahman Hassanein$^3$}\email[]{a.h.hassanein@tudelft.nl}
\author{Woutijn J. Baars$^3$}\email[]{w.j.baars@tudelft.nl}

\affiliation{$^1$Department of Mechanical Engineering, The University of Melbourne, Parkville, VIC 3010, Australia\\
$^2$School of Engineering, RMIT University, Melbourne, VIC 3000, Australia\\
$^3$Faculty of Aerospace Engineering, Delft University of Technology, 2629HS Delft,
The Netherlands}

\date{\today}

\begin{abstract}
This study shows that the turbulent velocities most strongly correlated with outer-scaled ($\delta$-scaled) wall-pressure fluctuations beneath a zero-pressure-gradient boundary layer reside within the logarithmic region. Even though contributions from the wake region are present, they are found to be statistically less dominant than those from the logarithmic region. The findings are based on bespoke measurements using an array of 63 microphones spanning 5$\delta$ in the streamwise direction (where $\delta$ is the boundary layer thickness), which synchronously captures space-time $p_w$ data alongside streamwise velocity fluctuations ($u$) from a single hotwire probe at the array's downstream end. The array is designed to spatially filter $p_w$ signals to uncover outer-scale contributions, by accurately resolving the large-scale portion of the frequency-wavenumber $p_w$ spectrum while avoiding aliasing of small-scale energy. This design, and its effectiveness in anti-aliasing, is validated against previously published low-Reynolds-number simulation datasets of turbulent boundary layer flow. Present experiments span a friction Reynolds number range of $1400 \lesssim Re_{\tau} \lesssim 5200$, over which the large-scale energy in the boundary layer grows significantly. This growth is reflected in both the frequency-wavenumber $p_w$ spectrum and the space-time $p_w$ correlations, both of which show scaling trends reflective of the large-scale pressure field convecting at an outer-scaled velocity of $0.75U_\infty$, where $U_\infty$ is the freestream velocity. The linear coherence between streamwise velocity and large-scale $p_w$ is directly quantified through space-time $p_w$--$u$ correlations, which show increasing magnitudes across the inner region with rising $Re_{\tau}$. At the top of the logarithmic region, the correlation contours resemble outer-scaled coherent structures akin to large- and very-large-scale motions. A clear Reynolds number trend is also evident in the average convection velocities inferred from $p_w$--$u$ correlations, which increasingly deviate from the local mean velocity towards the outer-scaled convection velocity across the inner region. These insights provide a critical foundation for leveraging wall-pressure fluctuations in modeling and control of high-$Re_{\tau}$ boundary layers, where large-scale motions increasingly dominate the turbulence dynamics.
\end{abstract}

\maketitle


\section{Introduction and motivation}
\label{intro}
Wall-pressure fluctuations ($p_w$) are a primary source of flow-induced noise in various wall-bounded turbulent flows, notable examples include the noise transmission into the aircraft cabin from the turbulent boundary layer (TBL) forming over its fuselage \citep{willmarth1975}, or aeroacoustic noise generated by pressure fluctuations scattering sound as they move past trailing edges \citep{lee2021}.
These fluctuations have been studied from a fundamental perspective for more than seven decades \citep{willmarth1975,willmarth1956}. 
Most investigations have adopted an experimental approach, typically involving wall-mounted microphones to measure $p_w$ imposed by the TBL grazing over (\emph{i.e.}, interacting with) them. 
The fast sampling capability of microphones, combined with the ability to vary the spatial offset between sensors, has enabled detailed statistical analyses through the computation of space-time plots/correlations of $p_w$ \citep{willmarth1962,bull1967,wallace2014,agastya2023}, or its frequency-wavenumber spectrum \citep{wills1964,abraham1998,damani2025,butt2026}.
The latter, in particular, provides critical input for models of turbulence-induced noise \citep{corcos1964,chase1980,yang2022}, which is classically attributed to structural vibrations excited by the turbulence-induced wall-pressure field.
Throughout this manuscript, $x$, $y$, $z$, and $t$ will indicate the streamwise, wall-normal, spanwise directions and time, respectively, with $k_x$, $k_z$ and $f$ representing streamwise and spanwise wavenumbers, and frequency, respectively.
The friction Reynolds number of the TBL is defined as $Re_{\tau} \equiv {\delta}{U_{\tau}}/{\nu}$, where $\delta$ is the TBL thickness (\emph{i.e.}, the outer length scale) and $\nu / U_\tau$ is the viscous or inner length scale.
Freestream velocity is denoted by $U_{\infty}$ and is the outer velocity scale, while the friction velocity, $U_{\tau} \equiv \sqrt{{\tau_w}/{\rho}}$, is the inner velocity scale. 
Here, $\tau_w$, $\rho$ and $\nu$ represent the mean wall shear stress, fluid density and kinematic viscosity, respectively.
Normalization by inner units will be represented by superscript `+'.

Wall-pressure fluctuations beneath a TBL are a spatially nonlocal function of the full source field in the pressure-Poisson equation \citep{tsuji2007,chang1999}, and can be obtained from a Green’s-function convolution of the source terms across the semi-infinite domain (\emph{i.e.}, above the wall).
These source terms underpin the dependence of $p_w$ on velocity fluctuations throughout the full flow domain \citep{panton1974,willmarth1975}.
This theoretical link between $p_w$ and the overlying velocity field was leveraged by \citet{farabee1991} and \citet{tsuji2007} to test classical scaling arguments for the $p_w$ frequency spectrum, which were inspired by scalings exhibited by the velocity spectra \citep{perry1986}. 
Their analysis was conducted for TBLs at low-to-moderate friction Reynolds numbers ($Re_{\tau}$ $\lesssim$ 6000), and revealed that the low-, mid-, and high-frequency domains of the $p_w$ spectrum scale with outer, intermediate, and Kolmogorov scaling parameters, respectively.
By comparing this against the spectral scaling exhibited by the velocity in different regions of the TBL \citep{perry1986}, it was inferred that the low-, mid- and high-frequency contributions to $p_w$ arise predominantly from the outer, overlap (\emph{i.e.}, logarithmic) and inner regions of the flow, respectively \citep{farabee1991,tsuji2007}.
However, previous studies did not focus on correlations of both streamwise and temporal scales, between $p_w$ and the turbulent velocities, at different $Re_{\tau}$ \citep{naka2015,gibeau2021}. Consequently, they could not directly associate space-time $p_w$ fluctuations with the $Re_{\tau}$-dependent behaviour of the coherent velocity structures in different regions of the TBL.
In this study, we report measurements that simultaneously acquire time series of wall-pressure and streamwise velocity across a TBL, and over large streamwise offsets. 
This focus on large scales stems from a persistent gap in understanding the extent to which large-scale velocity fluctuations---known to coexist throughout the BL \citep{marusic2010}---are correlated with the growth in large-scale energy of the wall-pressure fluctuations \citep{tsuji2007,klewicki2008,baars2024,deshpande2025}.
In this respect, considering the streamwise velocity component (as opposed to the wall-normal one) is deemed sufficient, as this component was shown to correlate with the wall-pressure at very low frequencies while the wall-normal velocity component did not \citep{baars2024,deshpande2025}.
We also clarify here that linear correlations between $p_w$ and isolated velocity signals can only be used to probe their statistical coupling, and cannot directly quantify the strength of $p_w$ `sources' in a strict Poisson-equation sense.

\begin{figure*}[tb!] 
\vspace{2pt}
\begin{tikzpicture}
   \node[anchor=south west,inner sep=0] (image) at (0,0) {
    \small
    \hspace{0.00cm}\includegraphics[width = 0.990\textwidth]{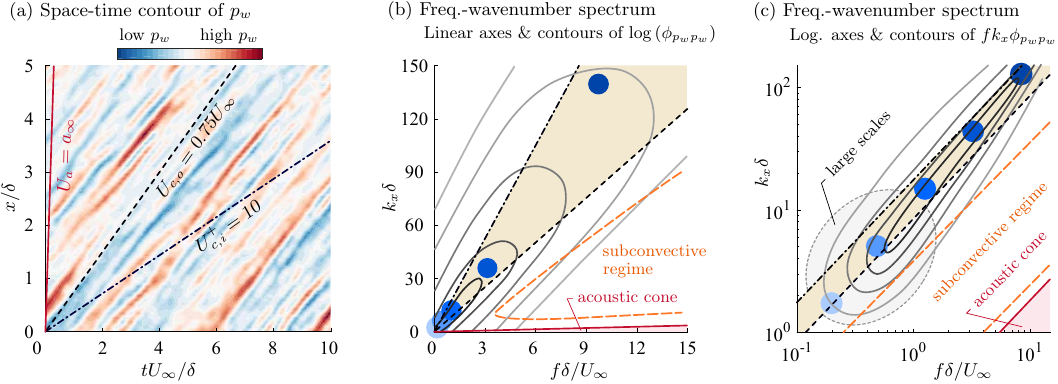} 
   };
   \begin{scope}[x={(image.south east)},y={(image.north west)}]
   \end{scope}
   \end{tikzpicture}
    \caption{(a) A space-time contour of wall-pressure fluctuations, $p_w$, with illustrative trend lines of a low ($U^+_{c,i} = 10$) and high ($U_{c,o} = 0.75U_\infty$) convection velocity. (b) Schematic illustration of the frequency-wavenumber spectrum of wall-pressure fluctuations, $\phi_{p_wp_w}(f,k_x)$, with linear axes; typically, logarithmically-spaced isocontours of $\phi_{p_wp_w}$ are considered. (c) Similar to subfigure (b) but now with logarithmic axes and linearly-spaced isocontours of the premultiplied spectrum, $fk_x\phi_{p_wp_w}$. The blue-shaded circles in subfigures (b) and (c) indicate the same five frequency-wavenumber scales.}
    \label{fig1}
\end{figure*}

Besides spectral scaling arguments, several past studies have attempted to infer the spatial origin of $p_w$-sources by investigating the scaling of their convection velocities ($U_c$).
This line of investigation dates back to the foundational work of Willmarth and co-workers \citep{willmarth1956,willmarth1962}, who experimentally reconstructed the space-time correlation of the wall-pressure, defined as:
\begin{equation}
\label{Corr2d_formula}
\begin{split}
{{\rho}_{{p_{w}}{p_{w}}}}({\Delta}x,{\Delta}t) = \: {R_{{p_{w}}{p_{w}}}}\bigg/\overline{p^2_w} \: = \big\langle \: {{p_w}(x,t)}\:{{p_w}(x+{\Delta}x,t+{\Delta}t)} \: \big\rangle \bigg/\overline{p^2_w},
\end{split}
\end{equation}
where ${\Delta}x$ and ${\Delta}t$ represent the streamwise and temporal offsets, respectively, and ${\big\langle}\cdot{\big\rangle}$ indicates ensemble averaging.
Per the approach of \citet{willmarth1975}, a convection velocity can be estimated from the slope of ${\rho}_{{p_w}{p_w}}$-isocontours. 
However, this estimate is representative of an \emph{average} convection velocity ($U_{c,a}$) that is an accumulation of all turbulent scales contributing to a given correlation level. To illustrate this, Fig.~\ref{fig1}(a) visualizes a sample of spatio-temporal wall-pressure data. This wall-pressure field is dispersive, in that fluctuations of $p_w$ associated with different statistical length scales of wall-pressure sources within the grazing turbulent flow convect at different speeds. Generally, large-scale (low-frequency) fluctuations convect faster than small-scale (high-frequency) ones and this type of analysis (as in Fig.~\ref{fig1}a) masks the scale-dependent behavior. As a consequence, several later studies employed a cross-spectral analysis of $p_w$ time series, measured for different streamwise offsets ($\Delta x$), to estimate convection velocities as a function of frequency, $\omega = 2{\pi}f$ (where ${\omega}$ is the circular frequency) \citep{willmarth1962,farabee1991,abraham1998}.
However, these inferred velocities are still a resultant of all contributing streamwise wavenumbers to the cross-spectra, and are biased towards the spatial scales that are resolved [according to the spatial Nyquist criterion, $k_x < 2\pi/(2{\Delta}x)$].

A statistical alternative to the aforementioned analyses is the usage of frequency-wavenumber spectra of the wall-pressure (further referred to as the $f$--$k_x$ wall-pressure spectrum, $\phi_{p_wp_w}\left(f,k_x\right)$, given that the streamwise wavenumber is being considered). This approach preserves both the spatial and temporal scales, as the two-dimensional (2D) spectrum is obtained via a space-time Fourier transform of $R_{{p_w}{p_w}}$ following:
\begin{equation}
\label{Spectra2d_formula}
\begin{split}
{{\phi}_{{p_{w}}{p_{w}}}}({k_{x}},f) = \int \int_{-{\infty}}^{\infty} {R_{{p_{w}}{p_{w}}}}({\Delta}x,{\Delta}t) {e^{-{j({k_{x}}{\Delta}x + 2{\pi}{f}{\Delta}t)}}} \; d({\Delta}x)\:d({\Delta}t).
\end{split}
\end{equation}
The $f$--$k_x$ wall-pressure spectrum facilitates the assessment of different convection velocities of $p_w$-signals associated with different energetic scales [Figs.~\ref{fig1}(b,c)]. This characteristic, referred to as the dispersive nature of the spatio-temporal wall-pressure field, results in the broad distribution of energy around a \emph{convective ridge} that is schematically shown in Figs.~\ref{fig1}(b,c), indicated by the shaded area between an inner- ($U^+_{c,i} = 10$) and outer-scaled ($U_{c,o} = 0.75U_\infty$) convection velocity; see also the $f$--$k_x$ wall-pressure spectra presented in the literature \citep[][among others]{corcos1964,abraham1998,yang2022,damani2025,butt2026}. Note that a constant convection velocity $U_c$ adheres to a constant ratio of $2{\pi}f/k_x$ [constant slopes in Figs.~\ref{fig1}(b,c)], making $f$--$k_x$ spectra suitable for assessing convection velocities of wall-pressure fluctuations.

Early applications of $f$--$k_x$ analysis of the wall-pressure field include studies by \citet{wills1964} and \citet{bull1967}, with the latter identifying two distinct `families' of convecting wavenumber components within the convective ridge: (i) high-wavenumber components exhibiting inner-scaled convection velocities ($U_{c,i} \sim 10U_{\tau}$), and (ii) low-wavenumber components associated with outer-scaled convection velocities ($U_{c,o} \sim 0.7$--$0.8U_{\infty}$).
These observations have been supported by several studies based on experiments \citep{schewe1983,thomas1983,farabee1991,abraham1998,damani2025,butt2026} and direct numerical simulation (DNS) \citep{choi1990,anantharamu2020,yang2022}, reinforcing the interpretation that high- and low-wavenumber $p_w$ contributions primarily originate from the inner and outer regions of the TBL, respectively. The same methodology of using the $f$--$k_x$ wall-pressure spectrum to infer convection velocities has also been adopted for TBL momentum fluctuations \citep{del2009,lehew2011,de2015,wilczek2015}.

In addition to the energetic convective ridge, turbulent velocity fluctuations induce wall-pressure energy in a broad region of the $f,k_x$-domain associated with a large range of convection velocities. A substantial regime is the \emph{subconvective regime}, as illustrated in Figs.~\ref{fig1}(b,c). Energy in this regime often drives wall vibrations and acoustic emissions of vehicle structures, and comprises relatively weak (low amplitude) pressure disturbances, occurring at combinations of low wavenumbers and frequencies propagating faster than the eddy-convection velocities [see \citep{chase1987}; the region demarcated with the dashed orange lines in Figs.~\ref{fig1}(b,c)]. While the origin of the subconvective signature remains unclear and is not the focus of this study, recent work has suggested that these fluctuations may be associated with either the streamwise-elongated very-large-scale motions, or the eddying motions impinging on the wall at an angle \citep{damani2025}. Note that subconvective pressure fluctuations are generally analyzed using $f$--$k_x$ wall-pressure spectra that are plotted with linear-linear axes, and logarithmically-spaced contours of energy, following $\log\left(\phi_{p_wp_w}\right)$ [see Fig.~\ref{fig1}(b)]. This representation is sufficient for analyzing the low-amplitude energy in the subconvective regime residing at moderate frequencies, but compresses the energetic region at low wavenumbers and frequencies into a narrow region towards $(f,k_x) = (0,0)$. Hence, when assessing the full-range of energetic scales, a representation with two logarithmic axes [for both $f$ and $k_x$, see Fig.~\ref{fig1}(c)] has certain advantages. First of all, when plotting contours of the premultiplied wall-pressure spectrum, following $fk_x\phi_{p_wp_w}$, the energy below a fraction of the contour is proportional to its contribution to the wall-pressure intensity (according to Parseval's theorem). In addition, it is more convenient to assess how the full range of large-scale/low-frequency energy is positioned relative to trend lines of constant convection velocity, which is the primary focus here. 
Finally, it is worth noting that the pressure signature of grazing acoustic (sound) waves can be present in the spatio-temporal wall-pressure data (particularly when performing experiments in non-anechoic facilities). Since sound waves convect at the acoustic velocity ($a_\infty$), the corresponding energy resides within an \emph{acoustic cone} as the trace velocities of the acoustic waves across the wall are sonic and supersonic. Because the current work considers flows with very low freestream Mach numbers ($M_\infty < 0.06$), the energy in the acoustic cone is non-overlapping with the turbulence-related energy [the scales of the axes, convective ridge, and acoustic cone of the schematic impressions of the $f$--$k_x$ wall-pressure spectrum in Fig.~\ref{fig1}(c) is representative of our data].

\begin{figure*}[tb!] 
\vspace{2pt}
\begin{tikzpicture}
   \node[anchor=south west,inner sep=0] (image) at (0,0) {
    \small
    \hspace{0.00cm}\includegraphics[width = 0.990\textwidth]{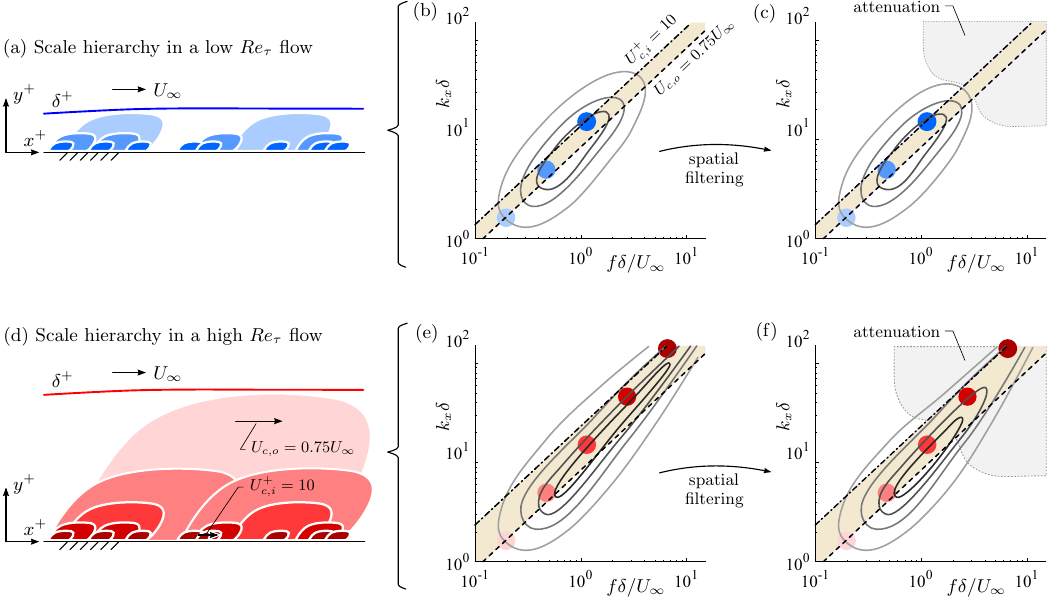} 
   };
   \begin{scope}[x={(image.south east)},y={(image.north west)}]
   \end{scope}
   \end{tikzpicture}
    \caption{(a) A conceptual representation of energy-containing scales in a zero-pressure gradient (ZPG) turbulent boundary layer at a relatively low Reynolds number. (b) Schematic of the premultiplied frequency-wavenumber spectrum of wall-pressure fluctuations, $fk_x\phi_{p_wp_w}$, with logarithmic axes. (c) Illustration of where the energy of the frequency-wavenumber spectrum is attenuated due to spatial filtering. (d,e,f) Similar to subfigures (a,b,c) but now for a relatively higher Reynolds number flow comprising a larger range of energy-containing scales.}
    \label{fig2}
\end{figure*}

Past simulation studies focusing on the convective ridge were mostly restricted to low $Re_{\tau}$ flows ($Re_\tau \lesssim 1000$), where the limited scale separation between the inner and outer-scaled turbulent motions complicates the task of distinguishing their respective contributions to the wall-pressure statistics. Figures~\ref{fig2}(a,b) schematically show a case of a low $Re_\tau$ flow with a small hierarchy of scales. The difference between the inner- and outer-scaled convection velocities is minor, and the convective ridge of the wall-pressure spectrum only spans a small portion of the $f,k_x$-domain. Moreover, the relative contribution of large scales to the $p_w$ energy are statistically weak at such values of $Re_{\tau}$ \citep{baars2024,deshpande2025}, making it difficult to identify scaling trends of, for instance, the convection velocity. Although some experimental studies have investigated moderately high-$Re_{\tau}$ flows, with an increase in scale-separation [Figs.~\ref{fig2}(d,e)], these flows also contain intermediate-scaled motions that impose distinct wall-pressure signatures. 
Their $p_w$-signatures convect at speeds falling between---and often overlapping with---the nominal inner- and outer-scaled bounds \citep{panton1974,luhar2014}. Nevertheless, in a relatively high $Re_\tau$ flow the difference between the inner- and outer-scaled convection velocity is larger, and the convective ridge of the wall-pressure spectrum spans a larger range in the $f,k_x$-domain. 
This, together with the log-log representation of the premultiplied $f$--$k_x$ wall-pressure spectrum, improves the assessment of the scale-specific $p_w$-signatures, which has not been done previously. 

Only a few detailed investigations in the literature have identified primary sources of small-scale wall-pressure fluctuations $p_w$ based on DNS at low Reynolds numbers of typically $Re_\tau \lesssim \mathcal{O}(10^3)$ \citep{chang1999,anantharamu2020,yang2022,agarwal2024}.
Some early experimental studies \citep{willmarth1963,thomas1983} also examined the correlations between $p_w$ and the overlying velocity field at low $Re_{\tau}$, uncovering the viscosity-dominated motions within the buffer layer to correlate most strongly with small-scale $p_w$.
On the other hand, no studies have explicitly assessed 
turbulent flow regions most strongly correlated with outer ($\delta$-
scaled) $p_w$ beneath TBLs, limiting our ability to (i) model and predict pressure fluctuations \citep{luhar2014,liu2020}, and (ii) use wall-pressure as input for wall-based flow control at high $Re_{\tau}$ conditions \citep{dacome:2025a}.
A key reason for this is the lack of prior work at sufficiently high Reynolds numbers (\emph{i.e.}, $Re_\tau \gtrsim 4000$), where the intermediate and outer-scaled motions begin to statistically dominate the flow dynamics over inner-scale ones \citep{marusic2010,deshpande2025} and drive the growth of the wall-pressure intensity \citep{klewicki2008,baars2024,deshpande2025,pirozzoli2025}.

\subsection{Novelty in the present approach}
Investigations of wall-pressure at sufficiently high $Re_\tau$ ($\gtrsim$ $\mathcal{O}$($10^3$)) are challenging due to measurement uncertainties related to spatial aliasing and the presence of facility noise contaminating low-frequency (large-scale) pressure fluctuations in the $p_w$ signals \citep{damani2025}. 
The present study introduces a dedicated experimental investigation and addresses these challenges by simultaneous measurements of $p_w$-signals along a long ($> 5\delta$) streamwise array of 63 microphones, to identify the turbulent flow regions most strongly correlated with outer ($\delta$) scaled $p_w$.
To avoid aliasing of small-scale energy onto the large scales, small-scale contributions are intentionally suppressed. 
This is achieved with sub-surface microphones that measure spatially-averaged wall-pressure fluctuations (following the approach by \citet{damani2025} and \citet{butt2026}) to filter out majoritarily the viscous-scaled $p_w$ energy, leading to an attenuated $p_w$ spectrum schematically shown in Figs.~\ref{fig2}(c,f).
Present experiments cover a range of $Re_\tau$ (1400 $\lesssim$ $Re_{\tau}$ $\lesssim$ 5200) that yields an increase in the intermediate and outer($\delta$)-scaled energy---from being statistically insignificant, to being dominant over small-scale contributions \citep{marusic2010,baars2024}.

Although past studies \citep{naka2015,gibeau2021} have explored space-time $p_w$--$u$ correlations across the TBL using long time series at a specific $x$-location (and relying on Taylor's hypothesis to infer spatial data), the streamwise array of the present study enables isolation of $p_w$--$u$ correlations influenced by motions that are large in both time and streamwise extent, without invoking Taylor's hypothesis.
This study starts by describing the experimental methodology (\S\,\ref{setup}), which includes a new streamwise-oriented microphone array for spatio-temporal measurements of wall-pressure. 
Data validation is presented in \S\,\ref{results1} and \S\,\ref{results2} by testing the $p_w$-spectra for: (i) the efficacy of the anti-aliasing methodology, and (ii) a collapse when outer-scaling is applied across a low-to-high $Re_{\tau}$ range \citep{dacome:2025a,pirozzoli2025}, respectively.
These wall-pressure data are then used in \S\,\ref{results3} for correlation analysis with the streamwise velocity data across the TBL, to infer the turbulent flow regions being coherent with the outer scaled $p_w$, their average convection velocity, and how these vary with Reynolds number.

\section{Experimental setup and methodologies}\label{setup}
\subsection{Turbulent boundary layer conditions and streamwise velocity measurements}
\begin{figure*}[tb!] 
\vspace{2pt}
\begin{tikzpicture}
   \node[anchor=south west,inner sep=0] (image) at (0,0) {
    \small
    \hspace{0.00cm}\includegraphics[width = 0.990\textwidth]{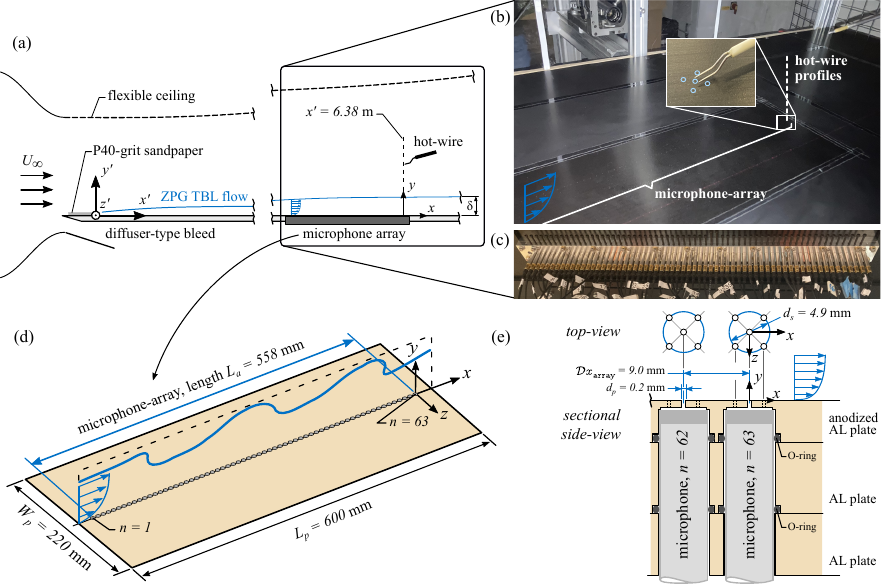} 
   };
   \begin{scope}[x={(image.south east)},y={(image.north west)}]
   \end{scope}
   \end{tikzpicture}
    \caption{(a) Setup for the TBL studies in the Delft University Boundary Layer Facility (DU-BLF, \citep{knoop:2025BLF}). (b) Photograph of the wind tunnel floor, including the panel that embeds the streamwise microphone array. Wall-normal profiles of hotwire measurements were performed above the most downstream wall-pressure measurement location. (c) Photograph of the streamwise microphone array, taken from the bottom: one can see the 63 microphones, and the two Aluminum (AL) plates and anodized AL top-plate housing the microphones. (d) Isometric view of the modular plate with the streamwise array of 63 microphones. (e) Detail of the two most downstream-located microphones: a top-view showing the arrangement of the five pinholes per $p_w$ measurement location, and a sectional side-view showing how the microphones are mounted beneath the top surface.}
    \label{fig3}
\end{figure*}

Experiments were carried out in the Delft University Boundary Layer Facility (DU-BLF) at the Delft University of Technology. A brief overview of the wind tunnel facility is provided here; further details are available in the literature \citep{knoop:2025BLF}. The tunnel is powered by a 75\,kW axial fan, and incorporates a heat-exchanger downstream. In the plenum, seven anti-turbulence screens are installed to achieve a low freestream turbulence intensity ($\mathrm{TI} \equiv \sqrt{\overline{uu}}/\overline{U}$ on the order of 0.1\,\%). The flow quality is further improved by an 11:1 area ratio contraction, comprising an outlet cross-sectional area of $0.9\times 0.6$\:m$^2$ (width $\times$ height). For studying TBL flow, a test section with a total streamwise length of 7.3\,m and a spanwise--wall-normal cross-section of 0.9\:$\times$\:0.5\:m$^2$ was installed. A new boundary layer developed on the suspended bottom wall with an elliptical leading edge; a diffuser-type bleed below the wall effectively bled the flow [see Fig.~\ref{fig3}(a)]. P40-grit sandpaper was applied on all four walls near the inlet to trip the boundary layer. The working section featured a flexible ceiling along its entire length, together with an over-pressure mesh, to achieve a streamwise ZPG development of the TBL.

A right-handed Cartesian coordinate system ($x,y,z$) is used throughout the paper, with the origin located at $x' = 6.38$\,m (where $x^\prime = 0$ corresponds to the downstream edge of the trip) and at the spanwise-center of the test section. At this nominal test location, wall-normal profiles of the streamwise velocity were acquired using hotwire anemometry (HWA). These acquisitions were synchronized with wall-pressure measurements from the streamwise microphone array (described in \S\,\ref{sec:pressmeas}). A TSI IFA-300 constant temperature anemometer (CTA) was employed, with a standard Dantec 55P15 boundary layer probe (comprising a sufficient length-to-diameter ratio of $l/d = 250$ \citep{hutchins2009hot}, with  $d =  5$\:\textmu m and $l = 1.25$\,mm). The analogue signal was low-pass filtered at 20\,kHz prior to A/D conversion using a NI\,DAQ\,PXI system (details in \S\,\ref{sec:pressmeas}). In-situ calibration was performed by fitting a fourth-order polynomial to 17 points of streamwise velocity; drifts in ambient temperature were corrected for following the method of \citet{hultmark2010temperature}. Wall-normal profiles of the streamwise velocity were acquired by employing a 300\,mm Zaber X-LRQ traversing system (10\:\textmu m step accuracy). 
For all four $Re_\tau$ test cases considered (achieved by way of varying the freestream velocity, $U_\infty$), a number of logarithmically-spaced $y$ locations were probed (see Table~\ref{tab1}). Parameters of the canonical TBL flow were inferred by fitting the measured mean velocity profiles to a composite velocity profile using log-law constants of $\kappa = 0.384$ and $B = 4.17$, with an added $\Delta y$ shift \citep{chauhan:2009a}. Table~\ref{tab1} summarizes the values of $U_\tau$, $\delta$ (based on composite profile), $Re_\tau$, and the acquisition length of the time series in terms of boundary layer turnovers ($T U_\infty/\delta$), for four nominal streamwise velocities: $U_\infty = 5$, 10, 15 and 20\,m/s, all at $x' = 6.38$\,m.

Figure~\ref{fig4}(a) shows wall-normal profiles of the streamwise velocity variance ($\overline{u^2}^+$), for all four testing conditions (\emph{i.e.}, four $Re_{\tau}$ test cases). 
The case with the lowest freestream velocity ($Re_{\tau} \approx 1400$) exhibits the highest spatial resolution, due to the smallest viscous-scaled hotwire sensing length ($l^+$; see Table~\ref{tab1}), and its $\overline{u^2}^+$ profile compares well with the one of a ZPG TBL DNS at a similar Reynolds number of $Re_{\tau} \approx 1300$ \citep{sillero2013}. For the relatively higher Reynolds number cases ($2800 \lesssim Re_{\tau} \lesssim 5200$), the increasing value of $l^+$ leads to a coarser spatial resolution. These measurements are compared with those of \citet{marusic_evolution_2015} at $Re_{\tau}$ $\approx$ 2800, 4300 and 5100, which were acquired using hotwire probes with a quasi-matched resolution ($l^+ \approx 24$). While the agreement is excellent in the outer region ($y^+$ $\gtrsim$ 100), a minor attenuation of $\overline{u^2}^+$ energy is observed in the inner region due to resolution limitations. This inner-region attenuation, however, does not impact the present work focusing on $p_w$--$u$ correlations associated with large ($\delta$-scaled) motions dominating the turbulence kinetic energy in the outer region \citep{marusic2010,baars2024,deshpande2025}.

\subsection{Fluctuating wall-pressure measurements}\label{sec:pressmeas}
\begin{figure}[tb!]
\includegraphics[width=1.0\textwidth]{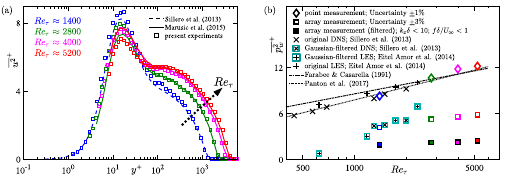} 
\caption{\label{fig4} 
(a) Wall-normal profile of the streamwise velocity variance, $\overline{u^2}^+$, and (b) variance of the wall-pressure fluctuations, $\overline{p^2_w}^+$, measured at various $Re_{\tau}$ in the present experiments. 
They are compared against previously published statistics represented by colored lines in (a), and by black symbols in (b) based on fully-resolved data. 
Also plotted in (b) are Gaussian-filtered versions of the published well-resolved statistics (in colored squares).
Dash-dotted line in (b) represents the empirical fit given by \citet{farabee1991}: $\overline{p^2_w}^+$ = 6.5 + 1.86\:ln\:($Re_{\tau}$/333), while the dotted line represents the empirical fit given by \citet{panton2017} based on the ZPG TBL data of \citet{schlatter2010}: $\overline{p^2_w}^+$ = 2.42\:ln($Re_{\tau}$) $-$ 8.96.}
\end{figure}

Before introducing the streamwise array of wall-pressure sensors, a dataset of resolved wall-pressure point measurements is briefly described. These data, acquired previously with a different setup \citep{knoop:2025BLF}, are used to validate wall-pressure intensity in the facility. Fluctuating pressure measurements were taken at $x^\prime \approx 6.38$\,m using multiple $\sfrac{1}{4}$\,in. GRAS\,46BE pressure microphones organized along the span (termed as `point' measurements henceforth). Five of those were mounted in sub-surface cavities, each of which communicated with the flow through a single-pinhole orifice with a diameter of $d = 0.4$\,mm (corresponding to $d^+ = 5$ to 17 for the four $U_{\infty}$ cases). Full details on data acquisition, the Helmholtz correction accounting for the sub-surface cavity resonance, and the removal of acoustic facility noise are available in published works \citep{baars2024,knoop:2025BLF}. Figure~\ref{fig4}(b) shows the wall-pressure variance ($\overline{p^2_w}^+$) from these measurements (open diamonds), for all four $Re_{\tau}$ test cases. 
They agree well with statistics from \citet{sillero2013} and \citet{eitel2014}---at similar spatial/grid resolution---and follow the empirical trend reported by \citet{farabee1991} and \citet{panton2017}. And so, Fig.~\ref{fig4}(b) confirms that $\overline{p^2_w}^+$ behaves as expected in the DUBLF.

\begin{table}[t!]
\begin{ruledtabular}
\begin{tabular}{cccccccccccc}
\multicolumn{11}{c}{(A) Wall-pressure array experiments in the Delft University Boundary Layer Facility:  \vspace{2mm}}\\
\hline
$U_{\infty}$ & $U_{\tau}$ & $\delta$ & $Re_{\tau}$ & $T{U_{\infty}}/{\delta}$ & Hotwire (HW) traverse & HW & Mic. & Mic. & ${\mathcal{D}}{x^+_{\rm array}}$ & ${\mathcal{D}}{t^+}$ & Color\\
(m/s) & (m/s) & (mm) &  &  &  & $l^+$ & $d^+_p$ & $d^+_s$ & &  $\times$ $10^{-2}$ & scheme\\
\hline
5 & 0.189 & 111 & 1400 & 19,500 & \#29 across: 1 $\lesssim$ $y^+$ $\lesssim$ $\delta^+$ & 12 & 2.4 & 58.8 & 108 & 4.5 & {\color{blue}Blue}\\
10 & 0.359 & 110 & 2800 & 20,000 & \#31 across: 3 $\lesssim$ $y^+$ $\lesssim$ $\delta^+$ & 24 & 4.8 & 117.6 & 216 & 16 & {\rd{Green}}\\
15 & 0.525 & 109 & 4000 & 20,000 & \#31 across: 5 $\lesssim$ $y^+$ $\lesssim$ $\delta^+$ & 33 & 6.7 & 164.15 & 300 & 35 & {\color{magenta}Magenta}\\  
20 & 0.684 & 108 & 5200 & 20,300 & \#31 across: 8 $\lesssim$ $y^+$ $\lesssim$ $\delta^+$ & 44 & 8.7 & 213.15 & 390 & 59 & {\color{red}Red}  \vspace{2mm}\\
\hline
\multicolumn{12}{c}{(B) DNS of ZPG TBL (\citet{sillero2013}):}  \vspace{2mm}\\
\hline
\multicolumn{2}{c}{$Re_{\tau}$} & \multicolumn{1}{c}{$({L_{x}}/{\delta})_{min}$} &  \multicolumn{1}{c}{$({L_{z}}/{\delta})_{min}$} & \multicolumn{1}{c}{${\mathcal{D}}{x^+}$} & \multicolumn{1}{c}{${\mathcal{D}}{z^+}$} & \multicolumn{1}{c}{${\mathcal{D}}{x^+_{\rm array}}$} & \multicolumn{5}{c}{cut-off ${k_x}{\delta}$ (for Gaussian filter)}  \\
\hline
\multicolumn{2}{c}{1300, 1500, 2000} & \multicolumn{1}{c}{$\sim$\:18} & \multicolumn{1}{c}{$\sim$\:6} & \multicolumn{1}{c}{7.0} & \multicolumn{1}{c}{4.07} & \multicolumn{1}{c}{108} & \multicolumn{5}{c}{10}   \vspace{2mm}\\
\hline
\multicolumn{12}{c}{(C) Well-resolved LES of ZPG TBL (\citet{eitel2014}):}   \vspace{2mm}\\
\hline
\multicolumn{2}{c}{$Re_{\tau}$} & \multicolumn{2}{c}{$T{U_{\infty}}$/$\delta$} & \multicolumn{1}{c}{$L_{z}$/$\delta$} & \multicolumn{1}{c}{${\mathcal{D}}{t^+}$} & \multicolumn{1}{c}{${\mathcal{D}}{z^+}$} & \multicolumn{5}{c}{cut-off ${f}{\delta}/{U_{\infty}}$ (for Gaussian filter)}\\
\hline
\multicolumn{2}{c}{1200, 1700, 2300} & \multicolumn{2}{c}{243} & \multicolumn{1}{c}{8} & \multicolumn{1}{c}{0.5} & \multicolumn{1}{c}{19.6} & \multicolumn{5}{c}{1}\\
\end{tabular}
\end{ruledtabular}
\caption{\label{tab1} Summary of the datasets used in the present study. Definitions of the various parameters can be found in either \S\,\ref{intro}, \S\,\ref{setup} or Fig.~\ref{fig3}. Note that the wall-pressure data of the current dataset (A) were measured synchronously with hotwire-based velocity measurements for all freestream velocities, except for $U_{\infty}$ = 15\,m/s.}
\end{table}

The present study reports new experiments using a streamwise array of 63 microphones (Fig.~\ref{fig3}) to capture spatial-temporal data of the fluctuating wall-pressure field. 
Each microphone in the streamwise array was mounted inside a circular sub-surface cavity, communicating with the boundary layer flow via multiple pinhole orifices. Multiple orifices spatially average the small-scale pressure fluctuations and avoid aliasing of high-wavenumber content onto larger scales, which we refer to here as anti-aliasing \citep[this approach is similar to the one taken by Damani \emph{et al.}][]{damani2025}. Five pinhole orifices of diameter $d_p = 0.2$\,mm were present per microphone (corresponding to $d^+_p = 2.4$ to 8.7 for the four $Re_\tau$ test cases, see Table~\ref{tab1}). 
Four orifices were arranged on a circle of diameter $d_s = 4.9$\,mm [Fig.~\ref{fig3}(e)], with the fifth at the center  (${d_s}/\delta$ $\approx$ 0.045; $d^+_s$ for different $Re_{\tau}$ reported in Table~\ref{tab1}). It was confirmed that the orifices did not affect the mean velocity profile (based on comparing hotwire measurements above the most downstream microphone to smooth-wall data). 
To capture the energetic spatial scales of wall-pressure, 63 microphones were arranged with an equidistant streamwise spacing of ${\mathcal{D}}{x_{\rm array}} = 9$\,mm, resulting in a total array length of $L_a = 558$\,mm [$\sim 5.1\delta$, see Figs.~\ref{fig3}(d,e)]. 
These array dimensions dictate the \emph{measurable} streamwise wavenumber range as: 1.2 $\lesssim$ ${k_x}{\delta}$ $\lesssim$ 38. 
This is however different from the wavenumber range `completely' and `partially' \emph{resolved} by the array, which is a function of the number, position ($d^+_s$) and size ($d^+_p$) of the pinholes and will be estimated directly based on the measured $p_w$-spectra later in $\S$III.

During plate design, the spacing between each microphone was dictated by the microphone-body diameter and the O-rings used to secure the microphones, which was kept to the minimum. 
As shown in the sectional side-view [Fig.~\ref{fig3}(e)] and the underside photograph of the array [Fig.~\ref{fig3}(c)], all microphones were inserted into an anodized aluminum (AL) top plate that housed the O-rings.
This assembly was closed using two additional AL bottom plates, housing another set of O-rings to ensure that the microphones were mounted straight and rigid.
The microphones used in this array were $\sfrac{1}{4}$\,in. GRAS\,40PH free-field pressure microphones, with a frequency response range of 5\,Hz to 20\,kHz and a $\pm 2$\,dB accuracy (and a $\pm 1$\,dB accuracy in the range of 50\,Hz up to 5\,kHz), a dynamic range of 32\,dBA to 135\,dB, and a sensitivity of 50\,mV/Pa. Here the reference pressure is taken as $p_{\mathrm{ref}} = 20$\:\textmu Pa. The sensor's dynamic range was verified to be adequate based on the expected levels of wall-pressure fluctuations [Fig.~\ref{fig4}(b)].
Although the streamwise arrangement of microphones results in the $Re_{\tau}$ and $\delta$ of the ZPG TBL increasing across the array length, their variation is within 10\% of the values reported at the array's downstream end, suggesting minimal influence on the present conclusions.

For data acquisition, several NI PXIe-4499 sound and vibration modules were used with the aid of a NI\,DAQ\,PXI system. These modules powered all 63 microphones via IEPE, and the voltage was filtered on-board prior to digitization with a 24-bit resolution; the output voltage of the hotwire anemometry system was acquired on an additional input channel. For the microphones, a calibration was applied to convert the voltage signals into units of Pascal. Microphone sensitivities were inferred through an in-situ calibration, performed with a GRAS\,42AG multifunction sound calibrator at a single frequency of $f = 250$\,Hz. The calibrator was equipped with a GRAS\,RA4800 adapter for proper positioning on the top surface. Regarding the data sampling, all 64 signals (from one HW and 63 microphones) were acquired synchronously at a rate of $f_s = 51.2$\,kHz and for the durations listed in Table~\ref{tab1}. Note that these durations correspond to \emph{one} position of the hotwire sensor, and that all wall-pressure statistics reported in this work are based on the full dataset, \emph{i.e.}, from the wall-pressure data corresponding to \emph{all} hotwire locations. For instance, for the lowest velocity setting of $U_\infty = 5$\,m/s, the total duration of wall-pressure time series is $TU_\infty/\delta \gtrsim 5.6 \cdot 10^5$ and is sufficient for converged large-scale spectral statistics \citep{damani2025,butt2026}.

\begin{figure*}[tb!] 
\vspace{2pt}
\begin{tikzpicture}
   \node[anchor=south west,inner sep=0] (image) at (0,0) {
    \small
    \hspace{0.00cm}
    \includegraphics[width = 0.990\textwidth]{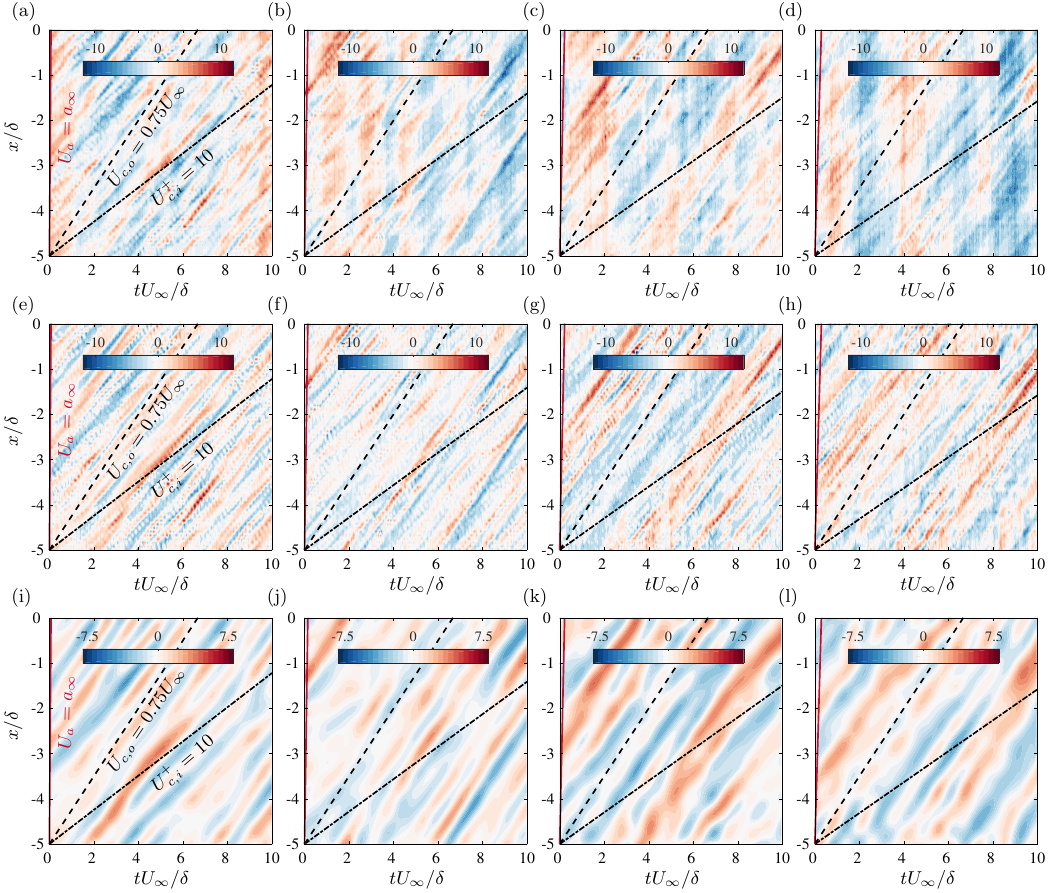}}; 
   \begin{scope}[x={(image.south east)},y={(image.north west)}]
   \end{scope}
   \end{tikzpicture}
    \caption{Space-time contour maps of the wall-pressure fluctuations, $p_w^+$, considering (a-d) the raw measured wall-pressure fluctuations, (e-h) the corrected wall-pressure fluctuations (following the two steps described in Appendix~A), and (i-l) the Fourier-filtered wall-pressure fluctuations ($f\delta/U_\infty < 1$ and $k_x\delta < 10$), for all four $Re_\tau$ test cases: (a,e,i) $Re_\tau \approx 1400$, (b,f,j) $Re_\tau \approx 2800$, (c,g,k) $Re_\tau \approx 4000$, and (d,h,l) $Re_\tau \approx 5200$. In all sub-plots, trend lines are included to indicate the inner- ($U^+_{c,i} = 10$, dash-dotted line) and outer-scaled ($U_{c,o} = 0.75U_\infty$, solid line) convection velocities.}
    \label{fig5}
\end{figure*}

Samples of measured (raw) wall-pressure data are shown as space-time contour maps in Figs.~\ref{fig5}(a-d), for all four $Re_{\tau}$ test cases. The raw data contain two erroneous contributions: (i) acoustic facility noise due to the non-anechoic tunnel environment, and (ii) viscous damping and resonance effects associated with the geometry of the pinholes and sub-surface cavity (forming a multi-pore Helmholtz resonator). All raw wall-pressure data were pre-processed to eliminate these erroneous contributions from the data before using them in any analyses. Both correction steps are described in Appendix~A, and are briefly reviewed here. First, for the acoustic facility noise, the space-time contour maps clearly show the acoustic component as fluctuations exciting almost all microphones simultaneously, due to the sonic and supersonic trace velocities of the acoustic waves along the array. Because the acoustic pressure fluctuations reside within the acoustic cone of the $f$--$k_x$ wall-pressure spectrum (as reviewed in \S\,\ref{intro}), they can be removed by setting the corresponding frequency/wavenumber components of the 2D Fourier transform of the pressure field to zero, \emph{e.g.,} the components for which $k_x < 2\pi f/a_\infty$ \citep{tinney2008}. Our space-time data allows for this approach, which is more effective than only subtracting the zero-valued spatial wavenumber or using reference sensors to subtract coherent acoustic signatures \citep[see approaches in the literature][]{farabee1991,naguib1996,richardson2023,baars2024}. Secondly, viscous damping and resonance effects associated with the multi-pore Helmholtz resonator geometries are not clearly visible in the space-time contour maps because of their frequency-dependence. These effects are accounted for by considering an empirical transfer kernel (inferred from a calibration experiment) relating the measured pressure to the pressure at the inlet of the pinholes. Corrected data are shown in Figs.~\ref{fig5}(e-h), showing the same data as were used for constructing Figs.~\ref{fig5}(a-d). 
Now, the signatures of acoustic noise are absent, and in all cases the (dispersive) convection signatures of the $p_w$ contours follow slopes that are predominantly bounded by the inner- ($U^+_{c,i} = 10$) and outer-scaled ($U_{c,o} = 0.75U_\infty$) convection velocity. 
We note here that the data plotted in Figs.~\ref{fig5}(e-h) comprises $p_w$-energy from scales that are fully-resolved (outer scales: $f\delta/U_\infty \lesssim 1$, 1.2 $\lesssim k_x\delta \lesssim 10$) as well as partially-resolved (intermediate-scales: $1 \lesssim f\delta/U_\infty \lesssim 4$, 10 $\lesssim k_x\delta \lesssim 38$), evidence of which will be provided based on the $p_w$-spectrum later in $\S$III. 
The latter is responsible for the relatively small-scale phenomena noted in Figs.~\ref{fig5}(e-h), which is distinct from the acoustic facility noise (given that has already been removed).
For completeness, Figs.~\ref{fig5}(i-l) depict the space-time contour maps for $p_w$-signals subjected to a low-pass Fourier-filter with bandpass regions of $f\delta/U_\infty \lesssim 1$ and 1.2 $\lesssim k_x\delta \lesssim 10$ (\emph{i.e.}, the fully-resolved scale range), where the lower-bound of $k_x$ is enforced by the maximum length of the streamwise array. 

With the resolved scales of the measurement in mind, the wall-pressure variance from the streamwise microphone array can be compared to the data from point measurements described earlier. 
Figure~\ref{fig4}(b) shows the current measurement data of $\overline{p^2_w}^+$ (open squares) and, while attenuated, the array data depicts the overall growth of energy with increasing $Re_{\tau}$. 
This growth nominally follows the same trend as the empirical fit of \citet{farabee1991}, with the only difference being a quasi-constant downward shift synonymous with the attenuation of viscous-scaled energy.
This is expected considering the $Re_{\tau}$-growth of the intermediate-scaled $p_w$-contributions \citep{klewicki2008,baars2024,deshpande2025,pirozzoli2025} that are (partially) resolved by the present multi-pinhole arrangement.
The consistency of the $Re_{\tau}$-growth of the $\overline{p^2_w}^+$ obtained from the array, with that exhibited by fully-resolved $\overline{p^2_w}^+$, is also indicative of the array spatial resolution decreasing only marginally for increasing $Re_{\tau}$ (relative to $Re_{\tau}$ $>$ 1400). 
The same Fig.~\ref{fig4}(b) also considers variances of the Fourier-filtered experimental data corresponding solely with the large (outer)-scaled contributions (with bandpass regions of $f\delta/U_\infty < 1$ and $k_x\delta < 10$), which show quasi-constant variation with $Re_{\tau}$ consistent with the findings of \citet{pirozzoli2025}.

\subsection{Published simulation datasets and associated analysis}\label{sec:publ}
While this paper mainly focuses on the analysis of new experimental data acquired with the streamwise microphone array, we also utilize two publicly available simulation datasets \citep{sillero2013,eitel2014} to validate trends seen in the former.
In particular, we assess the effect of spatial filtering and aliasing, \emph{i.e.}, issues that are inherent to the current experimental reconstruction of ${\rho}_{{p_w}{p_w}}$.
The first dataset is from a fully-resolved DNS of a ZPG TBL, published by \citet{sillero2013}.
We utilize their precomputed 2D streamwise-spanwise correlations of $p_w$, expressed as ${\rho}_{{p_w}{p_w}}({\Delta}x,{\Delta}z)$, at $Re_{\tau} \approx 1300$, 1500 and 2000.
While the streamwise (${L_x}/{\delta}$) and spanwise (${L_z}/{\delta}$) extents of ${\rho}_{{p_w}{p_w}}$ vary for the three $Re_{\tau}$ cases, Table~\ref{tab1} lists the minimum extents among those cases, which are sufficient for the present analysis.
The second dataset corresponds to a well-resolved large-eddy simulation (LES) of a ZPG TBL, published by \citet{eitel2014}. 
This LES was conducted at twice the grid spacing of a conventional DNS, along the streamwise and spanwise directions.
Although this resolution causes 10\% of the total turbulence kinetic energy dissipation to be unresolved, it is accounted for by adding a small body force which yields minimal differences when compared to a conventional DNS (see \citep{eitel2014} for further details).
The LES dataset comprises $p_w$ time series at selected $x$-locations across the entire spanwise extent of the domain (${L_z}$\:$\sim$ 8$\delta$), at a time resolution of ${\mathcal{D}}t^+$ $\lesssim$ 0.5 and for a duration of ${T}{U_{\infty}}/{\delta} \approx 243$. 
These data are used to compute space-time correlations, ${\rho}_{{p_w}{p_w}}({\Delta}z,{\Delta}t)$, at $x$-locations corresponding to $Re_{\tau} \approx$ 600, 1200, 1700 and 2300.

We mimic the spatial-filtering effect of the multiple pinholes in the microphone array by Gaussian-filtering the LES \citep{eitel2014} and DNS \citep{sillero2013} data (see Table \ref{tab1} and Appendix~B for further details) and comparing their $\overline{p^2_w}^+$ with array statistics in Figure~\ref{fig4}(b).
Both these Gaussian-filtered estimates exhibit $Re_{\tau}$-dependence similar to the array measurements, confirming the coexistence of both intermediate (which are $Re_{\tau}$-dependent) and $\delta$-scaled contributions to $\overline{p^2_w}^+$ \citep{pirozzoli2025} in the array data.
Note here that the comparison between the Gaussian-filtered LES/DNS and spatially-filtered array measurements is only qualitative, since the exact nature of filtering will be different in case of experiments.

Figures~\ref{fig12}(a,d) in Appendix~B show ${\rho}_{{p_w}{p_w}}({\Delta}x,{\Delta}z)$ and ${\rho}_{{p_w}{p_w}}({\Delta}z,{\Delta}t)$ at $Re_{\tau}$ $\approx$ 1300 and 1200, respectively, from the DNS and LES datasets. Here the smallest viscous-scaled streamwise offset achievable with the present microphone array (${\mathcal{D}}x^+_{\rm array} \sim 108$) at $Re_{\tau} \approx 1400$ (Table~\ref{tab1}) is also indicated. With the significantly large ${\mathcal{D}}x^+_{\rm array}$, only wavenumbers larger than $k_x > 2\pi/(2{\mathcal{D}}x^+_{\rm array})$ can be resolved with the experimental array. Without any measures to avoid aliasing, the unresolved high-wavenumber energy is projected onto low wavenumbers \citep{willmarth1975,abraham1998,damani2025}. How aliasing affects the one-dimensional (1D) autocorrelation ${\rho}_{{p_w}{p_w}}({\Delta}x)$, and how aliasing issues were avoided in the present experimental data, will be discussed in \S\,\ref{results}. We refer to Appendix~B for an evaluation of the effect of aliasing in the 2D streamwise-spanwise wavenumber spectrum of $p_w$.

\section{Results and discussion}
\label{results}
Results of the present experiments are organized into three subsections. First, \S\,\ref{results1} establishes the range of turbulent scales resolved in the present $p_w$ measurements and demonstrates the efficacy of the anti-aliasing strategy.
It also addresses the distinctive trends of the experimentally reconstructed space-time correlation, ${\rho}_{{p_w}{p_w}}$, and the 1D frequency and wavenumber spectra, ${\phi}_{{p_w}{p_w}}$ ($f$) and ${\phi}_{{p_w}{p_w}}$ ($k_x$), respectively.
Then, in \S\,\ref{results2}, $Re_{\tau}$-effects on the experimentally reconstructed ${\rho}_{{p_w}{p_w}}$ and ${\phi}_{{p_w}{p_w}}$ are discussed, specifically those associated with the growing influence of the large-scales in the overlying flow field.
Finally, \S\,\ref{results3} discusses the experimentally reconstructed space-time $p_w$--$u$ correlations. 

\begin{figure}[tb!]
\includegraphics[width=1.0\textwidth]{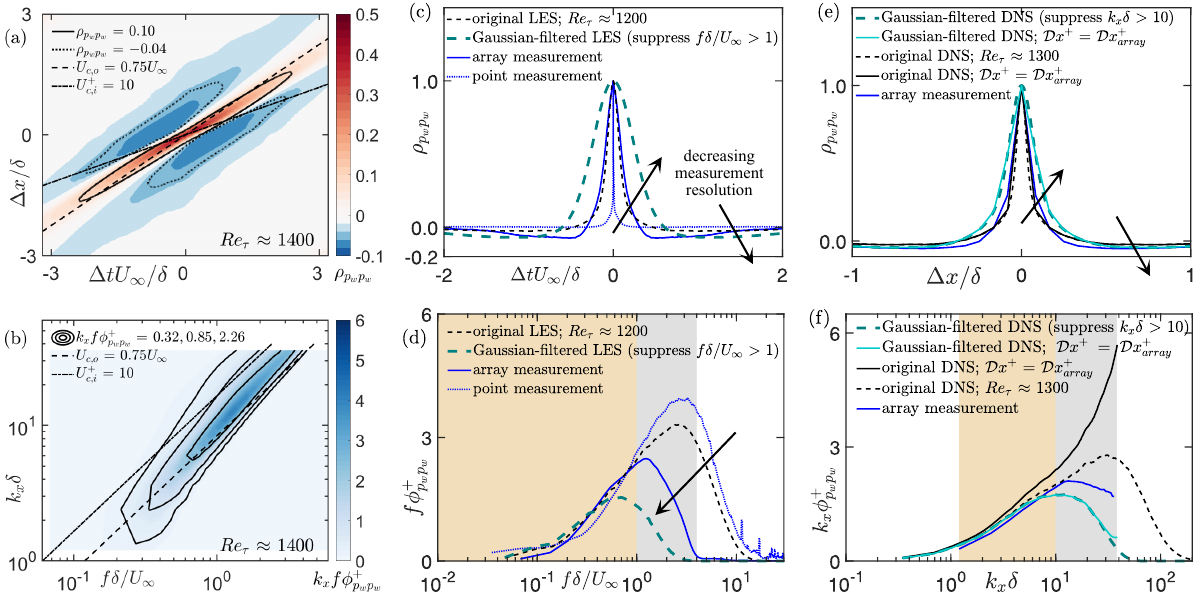} 
\caption{\label{fig6} 
(a) Space-time correlation coefficient, ${\rho}_{{p_w}{p_w}}$, and (b) premultiplied $f$--$k_x$ wall-pressure spectrum, $fk_x\phi_{p_wp_w}$, obtained from the present `array measurements' at $Re_{\tau}$ $\approx$ 1400.
Definitions of $U_{c,i}$ and $U_{c,o}$ are similar to those introduced previously in \S\,\ref{intro} and Fig.~\ref{fig1}. (c) Temporal correlation for ${\Delta}x$ = 0, and (e) spatial correlation for ${\Delta}t$ = 0, both extracted from subfigure (a) by horizontally and vertically slicing the space-time correlation coefficient, respectively. (d) Premultiplied frequency wall-pressure spectrum, and (f) premultiplied streamwise wavenumber wall-pressure spectrum, both obtained via integrating the $f$--$k_x$ wall-pressure spectrum in subfigure (b) over the full frequency and wavenumber ranges, respectively. These statistics in (c-f), which are measured via the microphone array, are compared against those obtained from well-resolved LES ($Re_{\tau}$ $\approx$ 1200) and resolved wall-pressure `point measurements' (in c,d), and the ZPG TBL DNS  (in e,f; $Re_{\tau}$ $\approx$ 1300). 
Subfigures (c-f) also include Gaussian-filtered estimates of ${\rho}_{{p_w}{p_w}}$ and ${\phi}^+_{{p_w}{p_w}}$ from the LES and DNS for qualitatively showcasing the effects of spatial filtering, while subfigures (e,f) consider estimates from the original as well as Gaussian-filtered DNS with ${\mathcal{D}}x^+$ = ${\mathcal{D}}x^+_{\rm array}$, to understand aliasing effects.
Background yellow and grey shading in subfigures (d,f) respectively correspond to the scale ranges for which the microphone array acquires fully-resolved (${f}{\delta}/{U_{\infty}}$ $\lesssim$ 1 and 1.2 $\lesssim$ ${k_x}{\delta}$ $\lesssim$ 10) and partially-resolved $p_w$-data (1 $\lesssim$ ${f}{\delta}/{U_{\infty}}$ $\lesssim$ 4 and 10 $\lesssim$ ${k_x}{\delta}$ $\lesssim$ 38).}
\end{figure}

\subsection{Resolved energy content of the measured wall-pressure fluctuations}
\label{results1}
To address the temporal and spatial scales resolved in the present $p_w$-measurements, we compare various second order statistics of the wall-pressure obtained from the experimental and numerical data.
Figure~\ref{fig6}(a) depicts the space-time correlation ${\rho}_{{p_w}{p_w}}$ estimated from the streamwise microphone array data at $Re_\tau \approx 1400$; reference lines representing the inner- ($U^+_{c,i}$ = 10) and outer-scaled ($U_{c,o}$ = 0.75$U_{\infty}$) convection velocities are also superimposed.
Isocontours at values of ${\rho}_{{p_w}{p_w}}$ = 0.10 and -0.04 are also plotted, with the positive correlation contour exhibiting a `wiggle' consistent with earlier observations \citep{willmarth1975}.
The wiggle arises due to a shift in the shallow slope of the correlation isocontours near ${\Delta}x$, ${\Delta}t$ $\sim$ 0, to a much steeper slope at large offsets.
This feature is representative of the differing convection velocities of small and large turbulent scales, which respectively influence the correlation at the small and large (${\Delta}x$, ${\Delta}t$) offsets.
The fact that large scales convect faster than small scales \citep{lehew2011,luhar2014,de2015,liu2020} explains the close alignment of the large- and small-scale ends of ${\rho}_{{p_w}{p_w}}$ with $U_{c,o}$ = 0.75$U_{\infty}$ and $U^+_{c,i}$ = 10, respectively.
In contrast, the negative correlation isocontours \citep{willmarth1962,willmarth1975} exhibit a slope closer to $U_{c,o}$ = 0.75$U_{\infty}$ for the full range of ${\Delta}x$ and ${\Delta}t$, reflecting the $p_w$-signatures synonymous with the large-scale motions.
This interpretation is also consistent with the fact that the two negative correlation isocontours are centered at $|{\Delta}x|$ $\sim$ $\delta$ and $|{\Delta}t{U_{\infty}}|$ $\sim$ $\delta$.
Further, the offset between these two negative correlation isocontours represents the average spatio-temporal spacing between two large-scale $p_w$ signatures of the same sign.
Overall, the alternating positive and negative values of ${\rho}_{{p_w}{p_w}}$ in both space and time is synonymous with the quasi-periodic organization of large-scale streamwise velocity fluctuations, as documented in previous experiments \citep[\emph{e.g.},][]{desilva2020}.

In order to understand how ${\rho}_{{p_w}{p_w}}$ is influenced by the spatially-filtered measurement of $p_w$ using the microphone array, Figs.~\ref{fig6}(c,e) plot ${\rho}_{{p_w}{p_w}}$(${\Delta}t{U_{\infty}}/{\delta}$, ${\Delta}x/{\delta}$ = 0) and ${\rho}_{{p_w}{p_w}}$(${\Delta}t{U_{\infty}}/{\delta}$ = 0, ${\Delta}x/{\delta}$), respectively, extracted from Fig.~\ref{fig6}(a).
Also included in Fig.~\ref{fig6}(c) is the temporal auto-correlation obtained from the resolved wall-pressure point measurements (recall \S\,\ref{sec:pressmeas}). 
These correlations are compared against ${\rho}_{{p_w}{p_w}}$ computed from the well-resolved LES and DNS at $Re_{\tau}$ $\approx$ 1200 and 1300, respectively, as shown in Figs.~\ref{fig6}(c,e). 
Besides these fully-resolved estimates, we also consider the Gaussian-filtered versions of ${\rho}_{{p_w}{p_w}}$ achieved by suppressing the energy residing at $f{\delta}/{U_{\infty}}$ $>$ 1 and ${k_x}\delta$ $>$ 10, respectively (see Appendix~B for details).
Evidently, the reduced measurement resolution results in a broadening of the positive peak in ${\rho}_{{p_w}{p_w}}$, and an increase in the magnitude of the negative peak (\emph{i.e.}, minima), with the latter also shifting to larger non-zero offsets.
All of these are well-known characteristics of correlations that are dominated by large-scale features \citep{wallace2014}, thereby aligning with the present intention behind conducting spatially-filtered $p_w$-measurements. 
With regards to interpretation of the flow physics, Figs.~\ref{fig6}(a,c,e) showcase the negative correlation contours to extend across 0.3 $\lesssim$ $|{\Delta}t{U_{\infty}}/{\delta}|$ $\lesssim$ 2 and 0.2 $\lesssim$ $|{\Delta}x/{\delta}|$ $\lesssim$ 2, which can be interpreted as the average extent of a large-scale $p_w$ signature.

The effect of spatial filtering is further evidenced by the premultiplied $f$--$k_x$ wall-pressure spectrum, plotted in Fig.~\ref{fig6}(b), alongside trend lines of inner- ($U^+_{c,i}$ $=$ 10) and outer-scaled ($U_{c,o}$ $=$ 0.75$U_{\infty}$) convection velocities.
The $f$--$k_x$ wall-pressure spectrum reveals an energetic convective ridge that more closely aligns with the outer than the inner convection velocity scaling.
This behavior is in stark contrast to the convective ridge of a fully-resolved $f$--$k_x$ wall-pressure spectrum at nominally similar $Re_{\tau}$, which typically aligns with the inner convection velocity scaling (see Fig.~3 of \citet{yang2022}). Integration of the $f$-$k_x$ spectrum along the entire $k_x$ or $f$ domain respectively yields the premultiplied frequency [Fig.~\ref{fig6}(d)] and streamwise wavenumber spectrum  [Fig.~\ref{fig6}(f)]. Both of them are compared in outer-scaling \citep{farabee1991,tsuji2007} against their well-resolved counterparts obtained from LES, DNS, as well as from the resolved wall-pressure point measurement conducted at comparable $Re_{\tau}$.
An attenuation of energy appears at high frequencies and streamwise wavenumbers for the $p_w$-spectra measured with the streamwise microphone array (relative to the other well-resolved estimates), while all the spectra collapse reasonably well at low frequencies and wavenumbers.
The fact that the former is an artifact of spatial filtering is reaffirmed by comparing with the 1D frequency (Fig.~\ref{fig6}d) and wavenumber (Fig.~\ref{fig6}f) spectra of the Gaussian-filtered $p_w$. 
The comparisons in Figs.~\ref{fig6}(d,f) suggest that the microphone array fully resolves the outer-scaled contributions to $p_w$ (based on $\delta$ and $U_{\infty}$), conforming to $f{\delta}/{U_{\infty}}$ $\lesssim$ 1 and 1.2 $\lesssim$ ${k_x}{\delta}$ $\lesssim$ 10 (indicated by yellow background shading), while partially resolving the intermediate scales:
1 $\lesssim$ ${f}{\delta}/{U_{\infty}}$ $\lesssim$ 4 and 10 $\lesssim$ ${k_x}{\delta}$ $\lesssim$ 38.
Here, except the ${k_x}\delta$ $\gtrsim$ 1.2 and ${k_x}\delta$ $\lesssim$ 38 bounds, none of the other bounds should be interpreted as a fixed $\delta$-scaled cut-off imposed by the setup, but rather as the nominal scale range for which the data are resolved ($\S$\ref{results2}).

As mentioned in \S\,\ref{setup}, the primary objective of conducting spatially-filtered $p_w$-measurement is to avoid aliasing of small-scale energy onto the large scales. 
We illustrate through Figs.~\ref{fig6}(e,f) that the array succeeds in this by considering ${\rho}_{{p_w}{p_w}}$ from the original as well as Gaussian-filtered DNS at a streamwise resolution similar to the lowest $Re_{\tau}$ dataset of the current experiment, \emph{i.e.}, ${\mathcal{D}}x^+$ $=$ ${\mathcal{D}}x^+_{\rm array}$ (referred as streamwise under-resolved version), and comparing them with their corresponding streamwise-resolved versions.
These scenarios respectively `simulate' cases where either single pinhole microphones (represented by original DNS), or multi-pinhole microphones (represented by Gaussian-filtered DNS), are spaced apart by ${\mathcal{D}}x^+_{\rm array}$.
The 1D wavenumber spectrum of $p_w$ obtained from both these `numerical experiments' are plotted in Fig.~\ref{fig6}(f); 
while the streamwise under-resolved case associated with original DNS manifests stark differences from its streamwise-resolved version, the one obtained from Gaussian-filtered spectrum exactly matches its streamwise-resolved version.
In case of the former, notably, aliased energy not only influences the $p_w$-spectrum near the intermediate scales (${k_x}{\delta}$ $\gtrsim$ 10) but also in the large-scale range (${k_x}{\delta}$ $<$ 10), highlighting aliasing as a significant source of error if fully-resolved $p_w$ are used to obtain ${\phi}_{{p_w}{p_w}}$.
On the other hand, consideration of the Gaussian-filtered ${p_w}$ significantly filters out the small-scales, eventually resulting in minimal errors owing to their aliasing.
This highlights the importance of the spatially-filtered $p_w$-measurements from the context of anti-aliasing, which permits accurate estimation of the low-wavenumber $p_w$-spectrum.
On a side note, Figs.~\ref{fig6}(b,f) indicate significantly low spectral energy at the smallest $k_x$ ($\sim$ 1.2$\delta$) resolved by the present experiments, suggesting that the present microphone array is sufficiently long for capturing the decay in large-scale (low-$k_x$) energy.

\begin{figure}[tb!]
\includegraphics[width=0.9\textwidth]{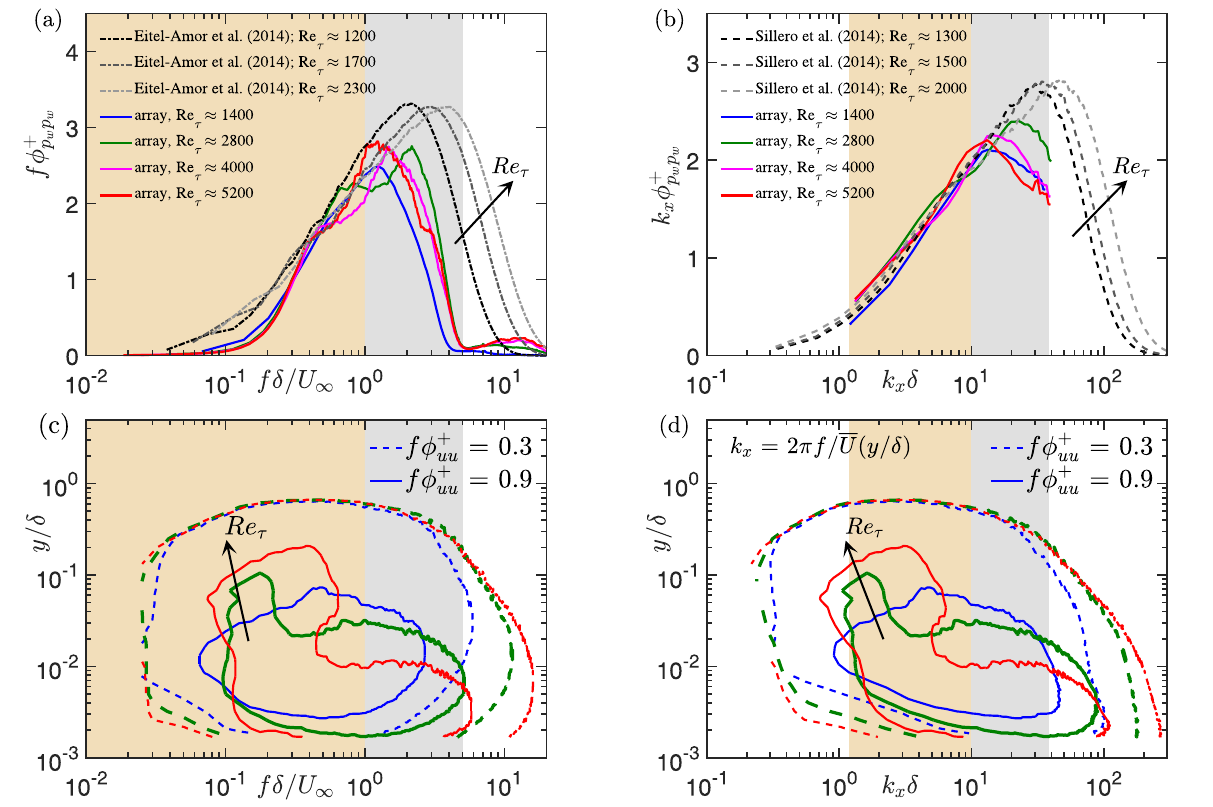} 
\caption{\label{fig7} 
(a) Premultiplied frequency and (b) streamwise wavenumber spectra of $p_w$ obtained by integrating the $f$--$k_x$ wall-pressure spectrum for all four $Re_\tau$ test cases. These spectra are compared against $p_w$-spectra inferred from the (a) well-resolved LES and (b) DNS data.
(c,d) Isocontours of premultiplied streamwise velocity spectra across the TBL as a function of frequency (in c) and streamwise wavenumbers (in d).
In (d), $k_x$ is obtained by imposing the Taylor's hypothesis.
Background yellow and grey shading in (a-d) respectively correspond to the scale ranges for which the microphone array acquires fully-resolved (${f}{\delta}/{U_{\infty}}$ $\lesssim$ 1 and 1.2 $\lesssim$ ${k_x}{\delta}$ $\lesssim$ 10) and partially-resolved $p_w$-data (1 $\lesssim$ ${f}{\delta}/{U_{\infty}}$ $\lesssim$ 4 and 10 $\lesssim$ ${k_x}{\delta}$ $\lesssim$ 38).}
\end{figure}
 
We next validate the microphone array measurements for the four $Re_\tau$ test cases considered.
Figures~\ref{fig7}(a,b) display the premultiplied frequency and streamwise wavenumber spectra of $p_w$, respectively, for the four $Re_{\tau}$ cases, each of which are obtained by integrating the corresponding $f$--$k_x$ wall-pressure spectrum [shown later in Fig.~\ref{fig8}(c)].
Also plotted for reference are the premultiplied $p_w$-spectra obtained from the well-resolved LES [1200 $\lesssim$ $Re_{\tau}$ $\lesssim$ 2300, in Fig.~\ref{fig7}(a)] and DNS [1300 $\lesssim$ $Re_{\tau}$ $\lesssim$ 2000, in Fig.~\ref{fig7}(b)].
Consistent with earlier observations in Figs.~\ref{fig6}(d,f), the experimental $p_w$-spectra are significantly attenuated at large $k_x$ and $f$, due to spatial filtering.
Further, the spectra collapse well with those from the LES/DNS at low frequencies and streamwise wavenumbers, as expected when applying the outer-scaling \citep{farabee1991,tsuji2007,dacome:2025a,pirozzoli2025}.
However, the exact $f$- and $k_x$-ranges associated with the outer-scaling vary with $Re_{\tau}$, owing to the competing effects of decreasing spatial resolution and $Re_{\tau}$-energization of the spectra (albeit marginal).
All these spectra in Figures~\ref{fig7}(a) also collapsed well on comparing with the experimentally obtained $p_w$-spectra by \citet{fritsch2022} for ZPG TBL (not shown here for brevity).
For this comparison with \citet{fritsch2022}, we only considered their experiments conducted with a zero airfoil angle of attack, and at the three upstream streamwise locations (to minimize upstream history effects), corresponding to 4000 $\lesssim$ $Re_{\tau}$ $\lesssim$ 8400.
As such, we can confirm from Figs.~\ref{fig7}(a,b) that the $p_w$-spectra are fully resolved within the ranges of: $f{\delta}/{U_{\infty}}$ $<$ 1 and 1.2 $\lesssim$ ${k_x}{\delta}$ $<$ 10, while being partially resolved in the intermediate scales: 1 $\lesssim$ ${f}{\delta}/{U_{\infty}}$ $\lesssim$ 4 and 10 $\lesssim$ ${k_x}{\delta}$ $\lesssim$ 38.

The close correspondence between this low frequency and low wavenumber range of $p_w$, and the energy-carrying large-scale velocity fluctuations, is assessed via the premultiplied spectrograms of the $u$-fluctuations in Fig.~\ref{fig7}(c) ($f{\phi}^+_{uu}$) and Fig.~\ref{fig7}(d) (${k_x}{\phi}^+_{uu}$) obtained from the present hotwire experiments. 
Note here that the streamwise wavenumber $u$-spectra are obtained by imposing Taylor's hypothesis, with a $y$-dependent mean velocity ($\overline{U}$($y$)), \emph{i.e.}, $k_x = 2\pi f /\overline{U}(y/\delta)$.
\citet{deshpande2023} have already demonstrated this to be a reasonably accurate assumption for estimating the streamwise velocity spectra in the logarithmic region of canonical boundary layers.
Figure ~\ref{fig7}(d) depicts two isocontours of the $u$-spectrogram for the three $Re_{\tau}$ cases at which hotwire profile measurements were conducted synchronously with $p_w$ (Table~\ref{tab1}).
The isocontours collectively indicate a clear $Re_{\tau}$-growth of the $u$-energy in the large-scale range ($f{\delta}/{U_{\infty}}$ $<$ 1 and ${k_x}{\delta}$ $<$ 10), which is also accurately resolved in the measured $p_w$-spectrum.
By contrast, the relatively small-scale range is known to be $Re_{\tau}$-invariant when scaled in viscous units \citep{hutchins2009hot,marusic2010,baars2020,gustenyov2025}).
Thus, Figures~\ref{fig7}(c,d) bring out the uniqueness of the present $Re_{\tau}$ range considered: the lower end ($\sim$ 1400) exhibits a relatively minor contribution of large-scale motions, while the higher end ($\sim$ 5200) includes a substantial large-scale contribution relative to the small-scale activity \citep{marusic2010}. These trends are expected to be reflected in the $p_w$--$u$ correlations, as discussed in \S\,\ref{results3}.

\subsection{Space-time correlations and $f$--$k_x$ wall-pressure spectra}
\label{results2}

\begin{figure}[tb!]
\includegraphics[width=0.85\textwidth]{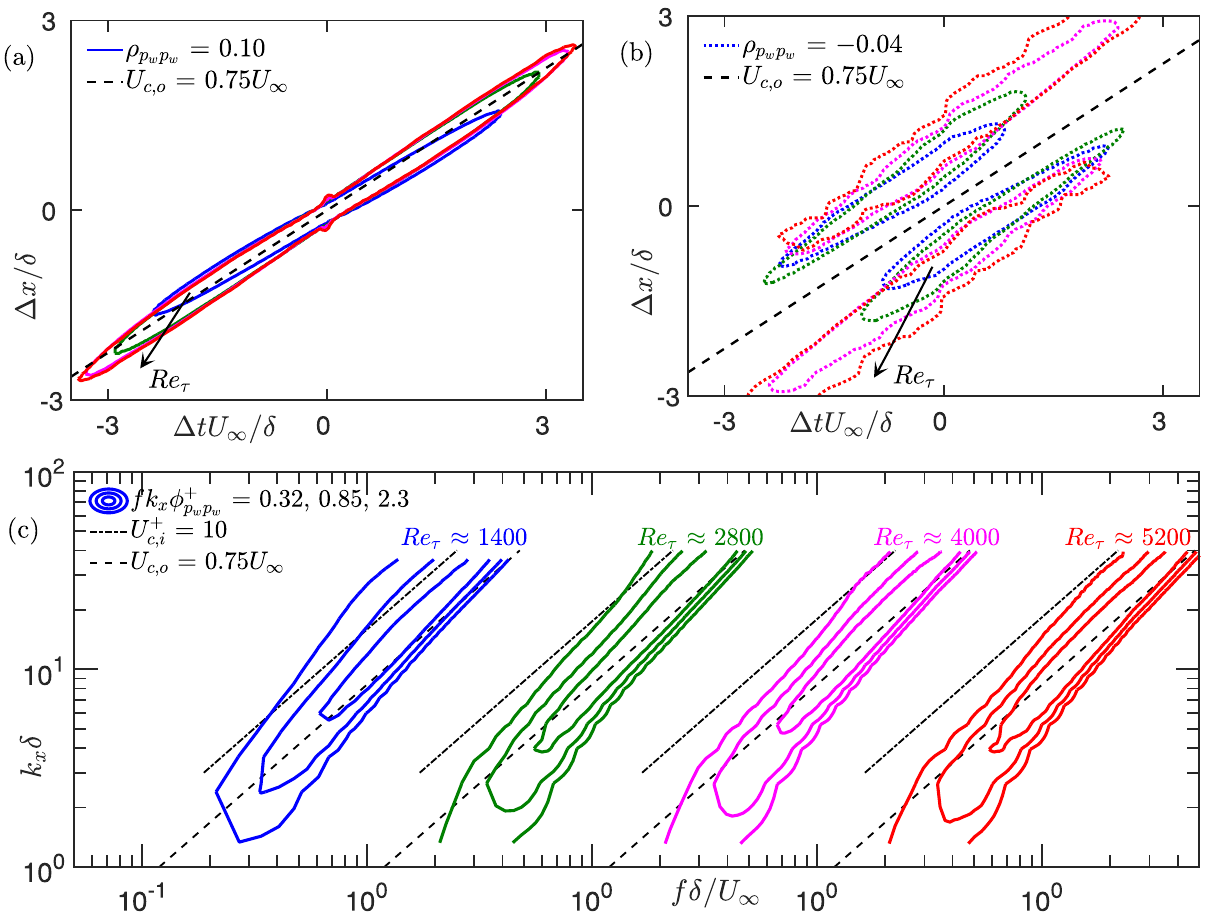} 
\caption{\label{fig8} 
Isocontours of the $p_w$ space-time correlation coefficient, corresponding to (a) 0.10 and (b) -0.04, reconstructed experimentally across 1400 $\lesssim$ $Re_{\tau}$ $\lesssim$ 5200.
(c) Contours of the premultiplied frequency-streamwise wavenumber spectra of $p_w$ obtained by computing the space-time Fourier transform of ${\rho}_{{p_w}{p_w}}$ across 1400 $\lesssim$ $Re_{\tau}$ $\lesssim$ 5200.
Note that the contours in subfigure (c) are consecutively offset in the horizontal direction with one decade of $f\delta/U_\infty$ for each increase in $Re_{\tau}$.
Definitions of $U_{c,i}$ and $U_{c,o}$ are similar to those introduced previously in \S\,\ref{intro} and Fig.~\ref{fig1}.}
\end{figure}

Trends in the space-time correlation (${\rho}_{{p_w}{p_w}}$) and the $f$--$k_x$ wall-pressure spectrum (${\phi}_{{p_w}{p_w}}$) with increasing values of $Re_\tau$ are driven by two effects: (i) the $Re_{\tau}$-energization of large-scale motions, and (ii) the decreasing spatial resolution and ability to resolve small scales, where the effect of latter has been argued to be very weak (see $\S$II. B).
Beginning with the analysis of ${\rho}_{{p_w}{p_w}}$, in Fig.~\ref{fig8}(a), isocontours at low values of ${\rho}_{{p_w}{p_w}}$ = 0.1 are plotted for all four $Re_{\tau}$ test cases (this low value makes these isocontours primarily representative of large turbulent scales). Also included for reference is the trend line of the outer-scaled convection velocity, $U_{c,o}$ = 0.75$U_{\infty}$. 
A significant shift in the slope of the positive ${\rho}_{{p_w}{p_w}}$ isocontour is observed alongside an increase in its spatio-temporal extent as $Re_{\tau}$ increases from 1400 to 2800, which is consistent with the energization of large-scale motions in the TBL flow [Figs.~\ref{fig7}(c,d)]. 
It must be noted here that the lowest $Re_{\tau}$ case is associated with insignificant large-scale energy, likely resulting in shorter time decorrelation than one would expect at relatively high $Re_{\tau}$.
While a further increase in $Re_{\tau}$ does not alter the slope of the isocontour, it does further expand the spatio-temporal extent of the positive ${\rho}_{{p_w}{p_w}}$ isocontours. 
Notably, the ridge of the ${\rho}_{{p_w}{p_w}}$ correlation aligns closely with $U_{c,o}$ = 0.75$U_{\infty}$ for the three highest $Re_{\tau}$ cases, suggesting they are influenced by a similar hierarchy of large scales.

When concentrating on the negative isocontours of ${\rho}_{{p_w}{p_w}}$, all cases show consistency with an outer-scaled convection velocity of $\sim$0.7--0.8$U_{\infty}$. 
The spatio-temporal extents of these isocontours are also found to increase significantly with increasing $Re_{\tau}$, and their centers gradually shift away from ${\Delta}x$, ${\Delta}t$ $\sim$ 0. 
This observation follows the previous discussion based on Figs.~\ref{fig6}(a-c), and further confirms the behaviour of the negative isocontours to be associated with the $Re_{\tau}$-energization of the large-scale motions in the TBL flow [Figs.~\ref{fig7}(c,d)].

Figure~\ref{fig8}(c) presents constant-energy isocontours of the premultiplied $f$--$k_x$ wall-pressure spectra for the four $Re_\tau$ test cases. All of them are computed via the space-time Fourier transform of the corresponding ${\rho}_{{p_w}{p_w}}$ using Eq.~(\ref{Spectra2d_formula}). The most energetic band of each $f$--$k_x$ wall-pressure spectrum corresponds to the convective ridge [see schematic in Fig.~\ref{fig1}(c)].
Here, we compare the spectral contours relative to both the inner- ($U^+_{c,i} = 10$) and outer-scaled ($U_{c,o} = 0.75U_\infty$) convection velocity scalings. At $Re_{\tau}$ $\approx$ 1400, the $p_w$-spectra contours are the same as those plotted previously in Fig.~\ref{fig6}(b), wherein the convective ridge already leans towards the $U_{c,o}$ scaling. As $Re_{\tau}$ increases, the convective ridge shifts further beyond the $U_{c,o}$ reference line. Meanwhile, the $U_{c,i}$ scaling gradually decreases in significance as it shifts beyond the convective ridge with increasing $Re_{\tau}$. 
This trend reflects the increasing large-scale contributions to $\overline{p^2_w}^+$ with increasing $Re_{\tau}$, in conjunction with a marginal drop in measurement spatial resolution.

\subsection{Wall-pressure$-$velocity correlations}\label{results3}

\begin{figure}[tb!]
\includegraphics[width=1.0\textwidth]{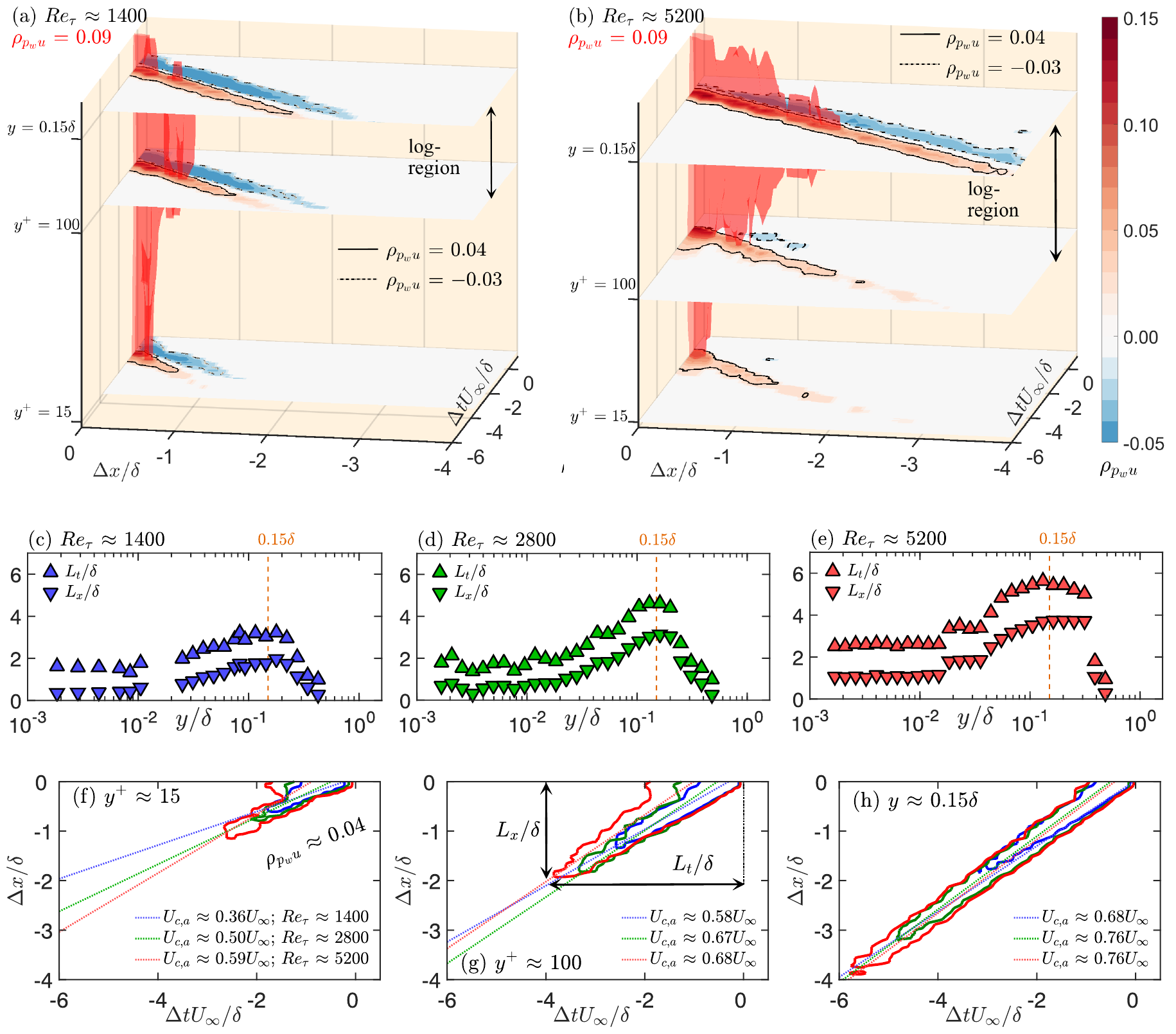} 
\caption{\label{fig9} 
(a,b) Isocontours (in red) of space-time wall-pressure$-$streamwise velocity correlations, ${\rho}_{{p_{w}}{u}}$ $=$ 0.04, across the TBL thickness at $Re_{\tau}$ $\approx$ (a) 1400 and (b) 5200.
Also plotted are `slices' of ${\rho}_{{p_{w}}{u}}$ at select wall-normal locations, which are represented via colored contours, solid (0.04) and dashed (-0.03) lines, respectively.
(c-e) plot the extents of ${\rho}_{{p_{w}}{u}}$ = 0.04 contour along time ($L_t$) and streamwise offset axis ($L_x$; defined in (g)) for $Re_{\tau}$ $\approx$ (c) 1400, (d) 2800 and (e) 5200.
It is noted that $L_x$ and $L_t$ are estimated solely based on the ridge of the main continuous isocontour that extends from ${\Delta}x$ = 0 and ${\Delta}t$ = 0, with their uncertainty dependent respectively on the streamwise ($\mathcal{D}x_{array}$) and temporal ($\mathcal{D}t$) resolution of the array and microphone.
(f-h) ${\rho}_{{p_{w}}{u}}$ = 0.04 isocontours compared for the three $Re_{\tau}$ at $y^+$ $\approx$ (f) 15, (g) 100 and (h) 0.15$\delta^+$.
$U_{c,a}$ is obtained by computing the slope of the ridge of the ${\rho}_{{p_{w}}{u}}$ contour, which is plotted in various colored dashed lines corresponding to different $Re_{\tau}$.}
\end{figure}

While wall-pressure--velocity correlations have been previously studied \citep{naka2015,gibeau2021,butt2026}, this subsection brings a novelty by cross-correlating $p_w$ with the streamwise velocity fluctuations ($u$) over large spatial (streamwise) and temporal offsets across 1400 $\lesssim$ $Re_{\tau}$ $\lesssim$ 5200. 
We define this space-time cross-correlation as:
\begin{equation}
\label{CrossCorr2d_formula}
\begin{split}
{{\rho}_{{p_{w}}{u}}}(y;{\Delta}x,{\Delta}t) =  \big\langle \: {{p_w}(x,t)}\:{{u}(y;x+{\Delta}x,t+{\Delta}t)} \: \big\rangle \bigg/{\sqrt{\overline{p^2_w}}\sqrt{\overline{u^2}(y)}},
\end{split}
\end{equation}
where 0 $<$ $y$ $\lesssim$ $\delta$ but ${\Delta}x$ = $x_{mic}$ - $x_{hw}$ $\le$ 0. 
This is because the hotwire probe is located at a fixed streamwise location of $x_{hw} = 0$ [directly above the most downstream microphone in the array, see Figs.~\ref{fig3}(a,b)], while all the microphones are placed upstream, \emph{e.g.}, $x_{mic}$ $\le$ 0.
While these $p_w$-signals are dominated by large-scale motions (due to spatial filtering), the $u$-signals used to compute ${{\rho}_{{p_{w}}{u}}}$ are unfiltered. 
Here, although the choice of using filtered $p_w$-signals (as opposed to the conventional fully-resolved $p_w$-signals \citep{naka2015,gibeau2021}) influences the ${{\rho}_{{p_{w}}{u}}}$ estimates quantitatively, it does not influence the qualitative trends of ${{\rho}_{{p_{w}}{u}}}$ for increasing $Re_{\tau}$ (which is the present focus).
This was confirmed by comparing ${{\rho}_{{p_{w}}{u}}}$ computed from the original as well as Gaussian-filtered $p_w$- and/or $u$-signals of the ZPG TBL LES dataset \citep{eitel2014}, results from which have not been presented here for purpose of brevity.

Figures~\ref{fig9}(a,b) show 3D isocontours of ${{\rho}_{{p_{w}}{u}}}$ = 0.09 obtained from array experiments, as a function of both spatial ($y$, ${\Delta}x$) and temporal offsets (${\Delta}t$), for the lowest ($Re_\tau \approx 1400$) and highest ($Re_\tau \approx 5200$) Reynolds numbers, respectively.
Also considered are 2D `slices' of ${{\rho}_{{p_{w}}{u}}}({\Delta}x,{\Delta}t)$ at three reference wall-normal locations of $y^+$ $\approx$ 15 (near-wall), $y^+$ $\approx$ 100 (lower bound of the logarithmic region), and $y$ $\approx$ 0.15$\delta$ (upper bound of the logarithmic region). Colored contours are used alongside solid and dashed lines corresponding to ${{\rho}_{{p_{w}}{u}}}$ = 0.04 and -0.03, respectively. While the red 3D isocontour is considered for gauging the variation of ${{\rho}_{{p_{w}}{u}}}$ across the TBL, the 2D slices are used for evaluating the $Re_{\tau}$-variation of ${{\rho}_{{p_{w}}{u}}}$ (as a function of $\Delta x$ and $\Delta t$) at specific wall-normal locations.

Starting with the low $Re_{\tau}$ case in Fig.~\ref{fig9}(a), the 3D ${{\rho}_{{p_{w}}{u}}}$ isocontour shows minimal variation across the near-wall and buffer regions (0 $<$ $y^+$ $<$ 50). However, near the lower bound of the logarithmic region (at $y^+ \approx 100$), the 3D isocontour extends farther in both ${\Delta}x$ and ${\Delta}t$; the same is confirmed by the corresponding 2D color contours at $y^+ \approx 100$.
Notably, these trends in the inner region are a consequence of the spatially-filtered measurement of $p_w$, which is effectively only resolving the large-scale contributions.
This filtering is crucial for unraveling the regions of the flow most strongly correlated with large-scale $p_w$, and avoids contamination by correlations associated with the small-scale $p_w$ [that are statistically dominant at $Re_{\tau} \lesssim \mathcal{O}(10^3)$, see Fig.~\ref{fig4}(b)]. In the relatively  high $Re_{\tau}$ case [Fig.~\ref{fig9}(b)], the qualitative behavior of both the 2D and 3D correlation contours remains similar, but the magnitude and spatial extent increase substantially compared to the low $Re_{\tau}$ case. 
Particularly, at $y$ $\approx$ 0.15$\delta$ the ${{\rho}_{{p_{w}}{u}}}$ = 0.04 2D contour extends beyond ${\Delta}x$ $\gtrsim$ 4$\delta$ and ${\Delta}t{U_{\infty}}$ $\gtrsim$ 6$\delta$. 
Considering that the present streamwise array can resolve correlations up to a maximum streamwise offset of ${\Delta}x$ $\sim$ 5.1$\delta$, these long isocontours are associated with the large outer-scaled motions whose signatures are consistent with VLSMs/superstructures reported in the literature \citep{marusic2010,deshpande2021,baars2024,deshpande2025}.
The present findings are consistent with previous studies describing the TBL outer region as the primary source of large-scale wall-pressure fluctuations \citep{bull1967,thomas1983,farabee1991,tsuji2007}, and provide quantitative support through identification of the linear coherent regions between $u$ and outer-scaled $p_w$.

To accentuate the variation in ${{\rho}_{{p_{w}}{u}}}$ across the Reynolds number range of 1400 $\lesssim$ $Re_{\tau}$ $\lesssim$ 5200, Figs.~\ref{fig9}(c-e) compare the spatial ($L_x$) and temporal ($L_t$) extents of the ${{\rho}_{{p_{w}}{u}}}$ = 0.04 2D contours at each $y$-location [see Fig.~\ref{fig9}(g) for definitions].
The choice of this low isocontour value was driven by our interest in analyzing $L_x$ and $L_t$ across the TBL thickness, where the correlation levels are weak very near the wall and towards the boundary layer edge. While the specific level influences the absolute values of $L_x$ and $L_t$, it does not affect their qualitative trends, ensuring the generality of the conclusions.
Both $L_x$ and $L_t$ are solely based on the ridge of the main continuous isocontour that extends from ${\Delta}x$ = 0 and ${\Delta}t$ = 0, with the ridge itself estimated through on a least squares fit.
Figures~\ref{fig9}(c-e) show that $L_x$ and $L_t$ exhibit a  plateau-type behavior in the near-wall region ($y$ $\lesssim$ 0.01$\delta$), followed by an increase with wall-normal distance throughout the logarithmic region, peaking near its upper bound ($y$ $\sim$ 0.15$\delta$), and then declining sharply as $y$ $\rightarrow$ $\delta$.
Importantly, the magnitudes of both $L_x$ and $L_t$ increase with $Re_{\tau}$, both near the wall and within the logarithmic region.
On the other hand, the correlation rapidly decays in the wake region ($y > 0.15\delta$), thereby suggesting that the logarithmic region exhibits the strongest and most spatially-extended correlation.
The trends associated with the logarithmic region are presumably related to the $Re_{\tau}$-increase in the hierarchy as well as energy of the large $\delta$-scaled eddying motions, which while being centered near the logarithmic region upper bound, also remain coherent down to the wall \citep{marusic2010,deshpande2021,baars2024,deshpande2025}.
Overall, the results in Fig.~\ref{fig9} indicate the dominance of large-scale $u$-$p_w$ correlations throughout the logarithmic region (100 $\lesssim$ $y^+$ $\lesssim$ 0.15$\delta^+$), with their influence strengthening as $Re_{\tau}$ increases. 
Their influence beyond the inner region is however dampened, consistent with the presence of a turbulent/non-turbulent interface (TNTI) near the boundary-layer edge \citep{lee2013,borrell2016}.

Figures~\ref{fig9}(f-h) compare the ${{\rho}_{{p_{w}}{u}}}$ = 0.04 isocontours for three $Re_\tau$ cases, at the same wall-normal locations considered in Figs.~\ref{fig9}(a,b).
The growth in correlation magnitude with increasing $Re_{\tau}$ is clearly evident. 
Particularly noteworthy is the change in the isocontour shape near the wall ($y^+$ $\approx$ 15) with increasing $Re_\tau$, whereas that of the isocontours at $y \approx 0.15\delta$ remain largely invariant with Reynolds number.
Quantitative differences in the isocontours are evaluated by estimating the slopes of contour ridges using a least-squares linear fit. These slopes represent the average convection velocity ($U_{c,a}$) of the eddying motions contributing to the $p_w$--$u$ correlations \citep{willmarth1975}, and are plotted as dashed lines in Figs.~\ref{fig9}(f-h).
A clear increase in $U_{c,a}$ is observed at $y^+ \approx 15$ with increasing $Re_{\tau}$, suggesting a growing influence of faster-convecting large scales in the near-wall region.
In contrast, at $y \approx 0.15\delta$, $U_{c,a}$ is in the order of 0.7--0.8$U_{\infty}$ and varies insignificantly with $Re_{\tau}$ [Fig.~\ref{fig1}(b)]. 
These results reaffirm the growth of linear coherence between $u$ and large-scale $p_w$ with increasing $Re_{\tau}$ across the TBL inner region.
Importantly, these conclusions are robust with respect to the choice of the isocontour level as the ridge slope (used to estimate $U_{c,a}$) is independent of that choice.

\begin{figure}[tb!]
\includegraphics[width=1.0\textwidth]{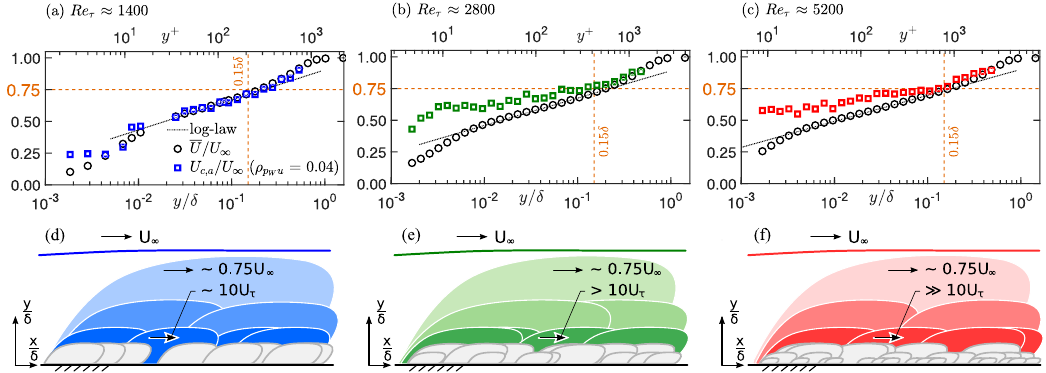} 
\caption{\label{fig10}
(a-c) Average convection velocity, $U_{c,a}$, estimated from the ridge of the ${\rho}_{{p_{w}}{u}}$ = 0.04 isocontour across the TBL thickness for $Re_{\tau}$ $\approx$ (a) 1400, (b) 2800, and (c) 5200.
$U_{c,a}$ is compared against the mean velocity profile, $\overline{U}$($y$), obtained directly from the hotwire measurements.
(d-f) Conceptual description of the turbulent eddying motions contributing to ${\rho}_{{p_{w}}{u}}$ for increasing $Re_{\tau}$, with increasing color intensity indicating stronger contribution to ${\rho}_{{p_{w}}{u}}$.}
\end{figure}

We now extend the analysis of $U_{c,a}$---estimated from the ${{\rho}_{{p_{w}}{u}}}$ contours---to consider its variation across the TBL.
Figures~\ref{fig10}(a-c) displays the convection velocity, $U_{c,a}$, as a function of wall-normal distance for three $Re_{\tau}$ test cases, alongside the mean velocity $(\overline{U})$ measured directly via hotwire anemometry. The trends of $U_{c,a}$ are explained using schematic representations of eddies in Figs.~\ref{fig10}(d-f). For the low Reynolds number ($Re_{\tau} \approx 1400$), both the small- and large-scale contributions are weak, due to spatial filtering and low $Re_{\tau}$ effects, respectively [Fig.~\ref{fig10}(d)].
Consequently, $U_{c,a}$ largely follows the mean velocity profile in Fig.~\ref{fig10}(a), with a slight bias towards $U^+_{c,i} \sim 10$ (or $\sim 0.3U_{\infty}$) in the very near-wall region.
However, an increase in $Re_{\tau}$ energizes the large- and intermediate-scale motions across the logarithmic and near-wall regions, while the small-scale contributions continue to be attenuated [Fig.~\ref{fig10}(e)].
This biases $U_{c,a}$ towards the convection velocity of large scales (\emph{i.e.}, $U_{c,o}$), resulting in the upward shift relative to the mean velocity profile in Fig.~\ref{fig10}(b).
The trend of an increasing $U_{c,a}$ with $Re_\tau$ continues for $Re_{\tau} \approx 5200$, in which there is an even stronger large-scale activity [Fig.~\ref{fig10}(f)] that further increases the estimated $U_{c,a}$ beyond the mean velocity profile in Fig.~\ref{fig10}(c).

Finally, it is worth reiterating that the quantitative trends in Figs.~\ref{fig9} and~\ref{fig10} are governed by the dominance of large ($\delta$)-scaled motions as captured by the spatially-filtered measurement of $p_w$. 
This is reaffirmed by the fact that the canonical convection velocity of the large scales ($U_{c,o}$ $\approx$ 0.75$U_{\infty}$; found here to be consistent with past studies \citep{bull1967,thomas1983,farabee1991,damani2025,butt2026}) closely matches $\overline{U}$ at $y$ $\approx$ 0.15$\delta$, which is where the large scale motions are also known to be centered \citep{marusic2010}.
In cases where $p_w$ is fully resolved, the correlation coefficient $\rho_{{p_w}u}$ (and its associated metrics such as the convection velocity estimate, $U_{c,a}$) are presumably biased by the small-scale contributions in the near-wall region, at least for the present low-to-moderate $Re_{\tau}$ cases.
But we have verified based on systematic analysis of the LES/DNS data \citep{eitel2014,sillero2013} that none of the present conclusions are an artifact of the $p_w$-filtering.

\section{Conclusions}\label{concl}
This study addressed a longstanding gap in our understanding of the turbulent flow regions strongly correlated with the outer ($\delta$) scaled $p_w$ in canonical TBLs.
While previous spectral analyses have inferred that low-, mid-, and high-frequency $p_w$-contributions arise predominantly from the outer, logarithmic, and inner regions of the TBL, the lack of direct space-time wall-pressure--velocity correlations has precluded an identification of the spatial locations dominating the coherence between $p_w$ and the overlying turbulent flow. 
Moreover, although outer-scaled $p_w$ signatures have been repeatedly observed in prior experiments and simulations, their isolation has remained challenging since their energetic signatures overlap with intermediate and inner scales---especially at moderate Reynolds numbers where scale separation is limited.
This work has overcome these limitations by conducting a dedicated experimental investigation spanning friction Reynolds numbers 1400 $\lesssim$ $Re_\tau$ $\lesssim$ 5200, so that the statistical dominance of large-scale motions progressively increases.
A streamwise array of 63 microphones spanning more than 5$\delta$ was used to capture space-time wall-pressure data, synchronously with streamwise velocity signals from a hotwire probe. 
Spectral and space-time correlation analyses confirm that the spatially-filtered data of the streamwise microphone array accurately captures the wall-pressure spectrum in the low frequency and wavenumber ranges, without relying on Taylor’s hypothesis. 
Comparison with LES and DNS-based `numerical experiments' further demonstrates that conventional microphone measurements with a single pinhole can lead to significant aliasing of small-scale energy onto the large-scale energy content---a pitfall avoided by the present design methodology of anti-aliasing.
These results firmly establish the array's efficacy in fully resolving the outer-scaled $p_w$ content within the scale ranges: $f\delta/U_\infty \lesssim 1$ and $1.2 \lesssim k_x\delta \lesssim 10$.

The space-time $p_w$-correlations reveal distinct structural signatures of large-scale motions, including alternating positive and negative lobes associated with the quasi-periodic nature of outer-layer turbulence. 
These correlations increase in spatial and temporal extent with Reynolds number, with negative contours exhibiting average convection velocities close to 0.75$U_\infty$ for all $Re_{\tau}$, representative of the outer-scaled $p_w$-signatures.
The $p_w$--$u$ correlations across the TBL further substantiate these trends. 
At relatively higher $Re_\tau$, these correlations grow in magnitude and spatial extent, particularly near the top of the logarithmic region ($y \approx 0.15\delta$), where contours are elongated over more than 4$\delta$ in streamwise separation and exceed 6$\delta$ in convective time scale---consistent with the signature of outer-scale motions synonymous with large- and very-large-scale motions/superstructures. 
This region is speculated to be the zone for the strongest linear coherence between large-scale $p_w$ and $u$ fluctuations.
The average convection velocities inferred from the ridge of $p_w$--$u$ correlations ($U_{c,a}$) exhibit a systematic upward shift with increasing $Re_\tau$ near the wall, reflecting the increasing influence of faster-convecting large scales on the near-wall flow. 
In contrast, $U_{c,a}$ remains nearly invariant with Reynolds number at $y \approx 0.15\delta$, aligning with the outer-scale convection velocity of $p_w$-signatures (0.7--0.8$U_\infty$). 
These trends reinforce the notion that large-scale correlations between $p_w$ and $u$ are most dominant across the logarithmic region and get stronger with increasing Reynolds number across the inner layer, while contributions from the wake region are relatively weaker than those in the logarithmic region.
These findings offer a strong foundation for advancing predictive models and flow control strategies that leverage wall-pressure measurements in high-Reynolds-number wall turbulence, where the TBL dynamics and wall-pressure intensities may become increasingly dominated by large-scale motions.

\begin{acknowledgments}
R. Deshpande is grateful for financial support from the University of Melbourne's Postdoctoral Fellowship, as well as the Visiting Fellow's Scheme awarded by the Faculty of Engineering \& Information Technology (FEIT).
W.J. Baars and A. Hassanein acknowledge financial support of the European Office of Aerospace Research and Development (EOARD) of the U.S. Air Force Office of Scientific Research (AFOSR) under Award No. FA8655-22-1-7168.\\
\end{acknowledgments}

The authors made the following contributions:
\textbf{R.D.:} Writing $-$ original draft, Conceptualization, Investigation, Methodology, Validation, Formal analysis, Funding acquisition;
\textbf{A.H.:} Writing $-$ review \& editing, Investigation, Methodology, Validation;
\textbf{W.J.B.:} Writing $-$ review \& editing, Conceptualization, Methodology, Funding acquisition, Supervision.\\

The authors declare that they have no conflict of interest.

\section*{Appendix A: Correction steps applied to the measured wall-pressure data}\label{correct}
\begin{figure}[tb!]
\includegraphics[width=0.750\textwidth]{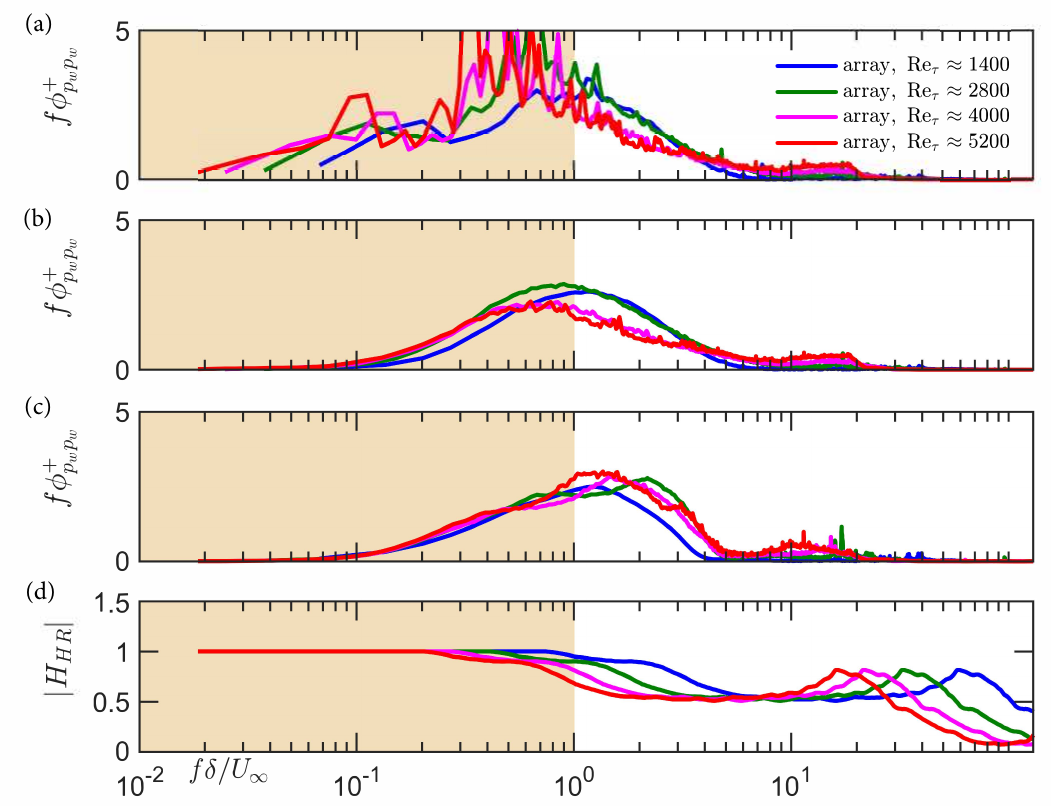} 
\caption{\label{fig11}
 (a) Premultiplied frequency spectra of the raw measured wall-pressure time series for the four $Re_\tau$ test cases (each spectrum is averaged from all 63 individual spectra corresponding to each microphone of the streamwise array). (b) Similar to subfigure (a) but now for the wall-pressure data after subtraction of the acoustic cone (so after applying correction step 1). 
 (c) Similar to subfigure (b) but now also with the Helmholtz correction applied (so after applying correction steps 1 and 2). (d) Gain of the Helmholtz transfer kernel, $\vert H_{HR} \vert$; four curves are shown corresponding to the four $Re_\tau$ test cases, although they are all identical and only offset in the horizontal direction due to the case-dependent normalization of the frequency-axis ($f\delta/U_\infty$).}
\end{figure}

Two correction steps were applied to the raw measured wall-pressure data (in units of Pascal), prior to using these data for all analyses. The two consecutive pre-processing steps are described here:\\[-14pt]
\begin{enumerate}[labelwidth=0.60cm,labelindent=0pt,leftmargin=1.00cm,label=(\roman*),align=left]
\item \noindent \textbf{Removing acoustic pressure fluctuations from the raw data:} $p_{\rm raw}(x,t) \rightarrow p_{\rm sub}(x,t)$. \\ Because of the non-anechoic tunnel environment, acoustic pressure fluctuations were removed from the measurement data. This was achieved by setting the frequency/wavenumber components of the 2D Fourier transform of the pressure field, corresponding to the passage of acoustic waves, to zero \citep{tinney2008}. First, the measured pressure field is 2D Fourier transformed, $P_{\rm raw}\left(k_x,f\right) = \mathcal{F}_{x,t}\left[p_{\rm raw}\left(x,t\right)\right]$, with $\mathcal{F}_{x,t}$ denoting the 2D Fourier transform in space and time. Hereafter, the frequency/wavenumber components with subsonic trace velocities along the spatial array are retained and inverse Fourier transformed to the physical domain: $p_{\rm sub} = \mathcal{F}^{-1}_{x,t}\left[P\left(k_x > 2\pi f/a_\infty,f < k_xa_\infty/(2\pi)\right)\right]$. This ensures that the frequency/wavenumber components with sonic and supersonic trace velocities (induced by the acoustic facility noise) are eliminated. Note that in this approach also the zero-valued streamwise wavenumber, $k_x = 0$, is eliminated. And thus, it is assumed that the hydrodynamic (turbulence-induced) wall-pressure fluctuations have negligible energy at streamwise scales longer than the array length ($\sim 5.1\delta$). This assumption is substantiated by the fact that at this scale ($k_x\delta \approx 2\pi/5.1 \approx 1.2$) the magnitude of the wall-pressure spectrum is already low according to the DNS spectra shown in Fig.~\ref{fig7}(b). To illustrate the removal of acoustic facility noise in the spectral domain, the frequency spectra of the measured (raw) wall-pressure data are shown in Fig.~\ref{fig11}(a), for the four $Re_\tau$ test cases. Spectral peaks indicate acoustic facility noise, consistent with the space-time contour maps in Figs.~\ref{fig5}(a-d). Spectra of the wall-pressure data after subtraction of the spatial mean wall-pressure are presented in Fig.~\ref{fig11}(b). Overall, distinct spectral peaks are absent in all spectra.\vspace{-6pt}
\item \noindent \textbf{Removing viscous damping and resonance effects:} $p_{\rm sub}(x,t) \rightarrow p(x,t)$. \\ Viscous damping and resonance effects occurred in measurements with the streamwise microphone array, as each sub-surface cavity and its pinholes form a multi-pore Helmholtz resonator. These effects were corrected using a transfer function, which was determined empirically using a calibration experiment with acoustic pressure excitation in an on-site anechoic facility \citep[][for details]{dacome2024}. No formulations for classical Helmholtz resonators could be used, because the cavity geometry was ill-defined due to the microphone’s fixed grid cap. The kernel is computed as $H_{HR}(f) = \langle P_m(f)P^*_i(f)\rangle/\langle P_i(f)P^*_i(f)\rangle$, where subscripts `m' and `i' stand for the measured pressure (at the microphone diaphragm) and pinhole-orifice inlet pressure (at the surface), respectively. The magnitude (gain) of $H_{HR}$ is shown in Fig.~\ref{fig11}(d) for the four $Re_\tau$ test cases; the curves are identical and only offset in the horizontal direction due to the case-dependent frequency normalization ($f\delta/U_\infty$). Resonance occurs near 3165\,Hz (around $f\delta/U_\infty = 1.8$ for $Re_\tau \approx 1400$), well above the energetic frequency scales of interest. Nevertheless, each wall-pressure signal after pre-processing step (i) was corrected by dividing its frequency-domain representation by the complex-valued kernel, to account for the frequency-dependent gain and phase relations between measured and inlet pressures. For one pressure signal at location $x_k$, the corrected signal in the spectral domain is $P(x_k,f) = \mathcal{F}_{t}\left[p_{\rm sub}(x_k,t)\right]/H_{HR}(f)$, and thus in the physical domain $p(x_k,t) = \mathcal{F}^{-1}_{t}\left[P(x_k,f)\right]$. When considering the spectra of Fig.~\ref{fig11}(b), their correction is equivalent to dividing them by the gain-squared, $\vert H_{HR}\vert^2$. The corrected spectra are shown in Fig.~\ref{fig11}(c) and are identical to those that were presented in Fig.~\ref{fig7}(a). At low frequencies, $\vert H_{HR}\vert \approx 1$ and the correction does not alter the spectra. At intermediate frequencies, the gain falls to $\sim 0.5$ and signifies that viscous damping dominates. Consequently, the correction intensifies the spectral levels here. Near the resonance frequency, the amplification due to resonance offsets the viscous damping and the gain reaches values near unity again. There the correction is again insignificant but also the wall-pressure spectra are less energetic in this frequency range.
\end{enumerate}

\section*{Appendix B: effects of aliasing and Gaussian-filtering on wall-pressure statistics}\label{sims}
\begin{figure}[tb!]
\includegraphics[width=1.0\textwidth]{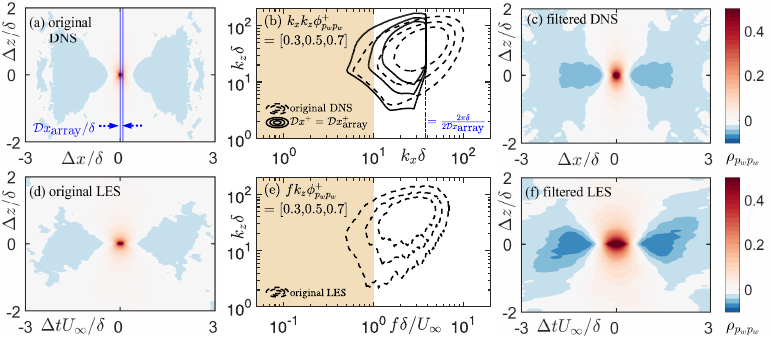} 
\caption{\label{fig12} 
(a) Streamwise-spanwise and (d) space-time correlation of $p_w$ obtained from ZPG TBL DNS \citep{sillero2013} and well-resolved LES \citep{eitel2014} at $Re_{\tau}$ $\approx$ 1300 and 1200, respectively.
Contours of premultiplied $p_w$-spectra as a function of (b) streamwise-spanwise wavenumbers and (e) frequency-streamwise wavenumbers, obtained by computing the Fourier transforms of ${\rho}_{{p_w}{p_w}}$ plotted in (a,d), respectively.
Also plotted in (b) is the $p_w$-spectrum computed from ${\rho}_{{p_w}{p_w}}$ based on ${\mathcal{D}}x^+$ = ${\mathcal{D}}x^+_{\rm array}$ (Table~\ref{tab1}), to evaluate aliasing effects.
(c) Streamwise-spanwise and (f) space-time correlation of $p_w$ obtained on inverse Fourier transforming the filtered $p_w$-spectra corresponding to (c) ${k_x}\delta$ $<$ 10 and (f) $f{\delta}/{U_{\infty}}$ $<$ 1 (indicated by background shading in (b,e), respectively).}
\end{figure}

This Appendix considers the published datasets (described in \S\,\ref{sec:publ}) to evaluate how the wall-pressure statistics are affected by aliasing and low-pass (spatial) filtering. Figures~\ref{fig12}(a,d) depict contours of ${\rho}_{{p_w}{p_w}}$(${\Delta}x$,${\Delta}z$) and ${\rho}_{{p_w}{p_w}}$(${\Delta}t$,${\Delta}z$) estimated from the DNS \citep{sillero2013} and well-resolved LES \citep{eitel2014} data at $Re_{\tau}$ $\approx$ 1300 and 1200, respectively.
While the 2D correlation along streamwise-spanwise direction [Fig.~\ref{fig12}(a)] was directly available online \citep{sillero2013}, ${\rho}_{{p_w}{p_w}}$(${\Delta}t$,${\Delta}z$) in Fig.~\ref{fig12}(d) has been computed from the space-time $p_w$ data by using (\ref{Corr2d_formula}), with ${\Delta}x$ being replaced by ${\Delta}z$.
Also plotted in Fig.~\ref{fig12}(a) for reference is the smallest streamwise offset achieved between two microphones (${\mathcal{D}}{x_{\rm array}}$) for the present array experiments at $Re_{\tau}$ $\approx$ 1400 (Table~\ref{tab1}). Negative correlation contours of ${\rho}_{{p_w}{p_w}}$ can be noted on either side of ${\Delta}x$ = 0 and ${\Delta}t$ = 0 in Figs.~\ref{fig12}(a,d); their magnitude levels however are significantly weak.

The dashed contours in Figs.~\ref{fig12}(b,e) respectively depict the wavenumber-wavenumber ($k_x$-$k_z$) and frequency-wavenumber ($f$--$k_z$) wall-pressure spectrum obtained by computing 2D and space-time Fourier transforms (Eq.~\ref{Spectra2d_formula}) of the original ${\rho}_{{p_w}{p_w}}$ in Figs.~\ref{fig12}(a,d). As expected based on the literature \citep{anantharamu2020,yang2022}, the energy is concentrated across the diagonal of the $k_x$--$k_z$/$f$--$k_z$ map, with small-scales (\emph{i.e.}, large $k_z$) corresponding with large $f$ and $k_x$, while large-scales (\emph{i.e.}, small $k_z$) corresponding with small $f$ and $k_x$.
If the 2D $p_w$-spectrum is computed based on ${\rho}_{{p_w}{p_w}}$ with smallest streamwise grid spacing equaling ${\mathcal{D}}{x^+_{\rm array}}$, then the corresponding energy contours are plotted with solid lines in Fig.~\ref{fig12}(b).
Not only does such a scenario result in the inability to resolve energy beyond $k_x > 2\pi/(2{{\mathcal{D}}x_{\rm array}})$, but this energy also gets aliased onto the relatively smaller wavenumbers.
As is evident in Fig.~\ref{fig12}(b), this aliasing issue can quantitatively influence almost the entire wavenumber range resolved in the array experiment, including the low-wavenumber end of the spectrum ($k_x\delta < 10$).
This highlights the need to filter out the small-scale energy of $p_w$, to avoid contamination of the low-wavenumber end of the spectrum (which is the primary focus of the study), which is achieved through spatially under-resolving the wall-pressure measured by each microphone (through using five pinholes) 

Besides avoiding the aliasing issue, spatial filtering of $p_w$ also increases the statistical contribution of large-scale wall-pressure fluctuations to ${\rho}_{{p_w}{p_w}}$ as well as $\phi_{{p_w}{p_w}}$.
This means that the space-time/2D maps associated with both these statistical quantities will be more representative of the large-scale $p_w$.
The effect of spatial/temporal filtering is simulated for the DNS and well-resolved LES data by considering only the spectral energy corresponding to ${k_x}{\delta}$ $<$ 10 and $f{\delta}/{U_\infty}$ $<$ 1 respectively (indicated through background yellow shading in Figs.~\ref{fig12}b,e) through a one-dimensional Gaussian filter.
Here, the Gaussian filter widths were based on these streamwise wavenumber and frequency cut-offs for the DNS and LES data, respectively. 
The corresponding 2D/space-time correlation ${\rho}_{{p_w}{p_w}}$ maps are plotted in Figs.~\ref{fig12}(c,f), which depict a statistically stronger negative correlation than that noted in the original ${\rho}_{{p_w}{p_w}}$ maps in Figs.~\ref{fig12}(a,d), consistent with present experimental observations (Fig.~\ref{fig8}b).

\bibliography{references}

\end{document}